\documentclass[
 reprint,
 amsmath,amssymb,superscriptaddress,
 aps
]{revtex4-2}

\usepackage{graphicx}
\usepackage{dcolumn}
\usepackage{bm}
\usepackage{amsthm}
\usepackage{subfigure}
\usepackage{caption}
\usepackage{overpic}
\usepackage{booktabs}
\usepackage{threeparttable}
\usepackage[inkscapelatex=false]{svg}
\usepackage{makecell}
\usepackage{xcolor}
\usepackage{hyperref}

\newtheorem{definition}{Definition}
\newtheorem{proposition}{Proposition}

\begin{document}

\title{Efficient and Error-Resilient Data Access Protocols for a Limited-Sized Quantum Random Access Memory}

\author{Zhao-Yun Chen}
 \email{chenzhaoyun@iai.ustc.edu.cn}
\author{Cheng Xue}
 \affiliation{Institute of Artificial Intelligence, Hefei Comprehensive National Science Center, Hefei, Anhui, 230088, P. R. China}
\author{Yun-Jie Wang}
 \affiliation{Institute of the Advanced Technology, University of Science and Technology of China, Hefei, Anhui, 230088, P. R. China}
\author{Tai-Ping Sun}
\author{Huan-Yu Liu}
 \affiliation{CAS Key Laboratory of Quantum Information, University of Science and Technology of China, Hefei, Anhui, 230026, P. R. China}
 \affiliation{CAS Center For Excellence in Quantum Information and Quantum Physics, University of Science and Technology of China, Hefei, Anhui, 230026, P. R. China}
  \affiliation{Hefei National Laboratory, Hefei, Anhui, 230088, P. R. China}
\author{Xi-Ning Zhuang} 
 \affiliation{CAS Key Laboratory of Quantum Information, University of Science and Technology of China, Hefei, Anhui, 230026, P. R. China}  
 \affiliation{Origin Quantum Computing Company Limited, Hefei, Anhui, 230026, P. R. China}
\author{Meng-Han Dou} 
\author{Tian-Rui Zou} 
\author{Yuan Fang} 
 \affiliation{Origin Quantum Computing Company Limited, Hefei, Anhui, 230026, P. R. China}
\author{Yu-Chun Wu}
 \email{wuyuchun@ustc.edu.cn}
 \affiliation{Institute of Artificial Intelligence, Hefei Comprehensive National Science Center, Hefei, Anhui, 230026, P. R. China}
 \affiliation{CAS Key Laboratory of Quantum Information, University of Science and Technology of China, Hefei, Anhui, 230026, P. R. China}
 \affiliation{CAS Center For Excellence in Quantum Information and Quantum Physics, University of Science and Technology of China, Hefei, Anhui, 230026, P. R. China}
 \affiliation{Hefei National Laboratory, Hefei, Anhui, 230088, P. R. China}
\author{Guo-Ping Guo}
  \email{gpguo@ustc.edu.cn} 
 \affiliation{Institute of Artificial Intelligence, Hefei Comprehensive National Science Center, Hefei, Anhui, 230026, P. R. China}
 \affiliation{CAS Key Laboratory of Quantum Information, University of Science and Technology of China, Hefei, Anhui, 230026, P. R. China}
 \affiliation{CAS Center For Excellence in Quantum Information and Quantum Physics, University of Science and Technology of China, Hefei, Anhui, 230026, P. R. China}
  \affiliation{Hefei National Laboratory, Hefei, Anhui, 230088, P. R. China}
 \affiliation{Origin Quantum Computing Company Limited, Hefei, Anhui, 230026, P. R. China}


\date{\today}

\begin{abstract}
Quantum Random Access Memory (QRAM) is a critical component for loading classical data into quantum computers. While constructing a practical QRAM presents several challenges, including the impracticality of an infinitely large QRAM size and a fully error-correction implementation, it is essential to consider a practical case where the QRAM has a limited size. In this work, we focus on the access of larger data sizes without keeping on increasing the size of the QRAM. Firstly, we address the challenge of word length, as real-world datasets typically have larger word lengths than the single-bit data that most previous studies have focused on. We propose a novel protocol for loading data with larger word lengths $k$ without increasing the number of QRAM levels $n$. By exploiting the parallelism in the data query process, our protocol achieves a time complexity of $O(n+k)$ and improves error scaling performance compared to existing approaches. Secondly, we provide a data-loading method for general-sized data access tasks when the number of data items exceeds $2^n$, which outperforms the existing hybrid QRAM+QROM architecture. Our method contributes to the development of time and error-optimized data access protocols for QRAM devices, reducing the qubit count and error requirements for QRAM implementation, and making it easier to construct practical QRAM devices with a limited number of physical qubits.
\end{abstract}

\maketitle
\section{Introduction}
Quantum computing has experienced rapid growth in recent years, with the potential to solve problems that classical computers cannot efficiently solve~\cite{quantumsupremacy2019, strongquantumsupremacy, factoring}, and to improve performance on computationally expensive classical tasks such as quantum machine learning~\cite{huang2021power, biamonte2017quantum, Kerenidis2020Quantum} and quantum differential equation solvers~\cite{differentialanalysis, homotopyperturbation, dissipative, QCFD}. Nevertheless, a major challenge in quantum computing resides in the classical data loading task~\cite{cerezo2022challenges, ciliberto2018quantum}. If $N$ data entries are encoded in the quantum circuit, the gate complexity is at least $O(N)$, which may negate any ``exponential'' quantum speedup~\cite{readthefineprint}. 

Given this challenge for numerous quantum algorithms, quantum random access memory~(QRAM) was introduced to address this problem~\cite{QRAM, QRAMArch, OpticalQRAM, QWQRAM2, QWQRAM, AtomQRAM, PhotonQRAM}. QRAM, a quantum analog of classical random access memory, can efficiently load classical data into a quantum computer and allows querying in quantum superposition. The QRAM achieves its efficiency by exploiting circuit-level parallelism, using $O(N)$ ancillary qubits to load $O(N)$ data items with $O(\log N)$ time complexity. The exponential qubit count required for implementing a scalable QRAM has been one of the most challenging aspects, as even a small error rate in qubits can lead to large errors in the entire system. To overcome this issue, a promising QRAM architecture called the bucket-brigade architecture was proposed in \cite{QRAMArch} which has already been demonstrated to be error-resilient \cite{ErrorResilience}.

\begin{figure*}
    \centering
    \includegraphics[width=\linewidth]{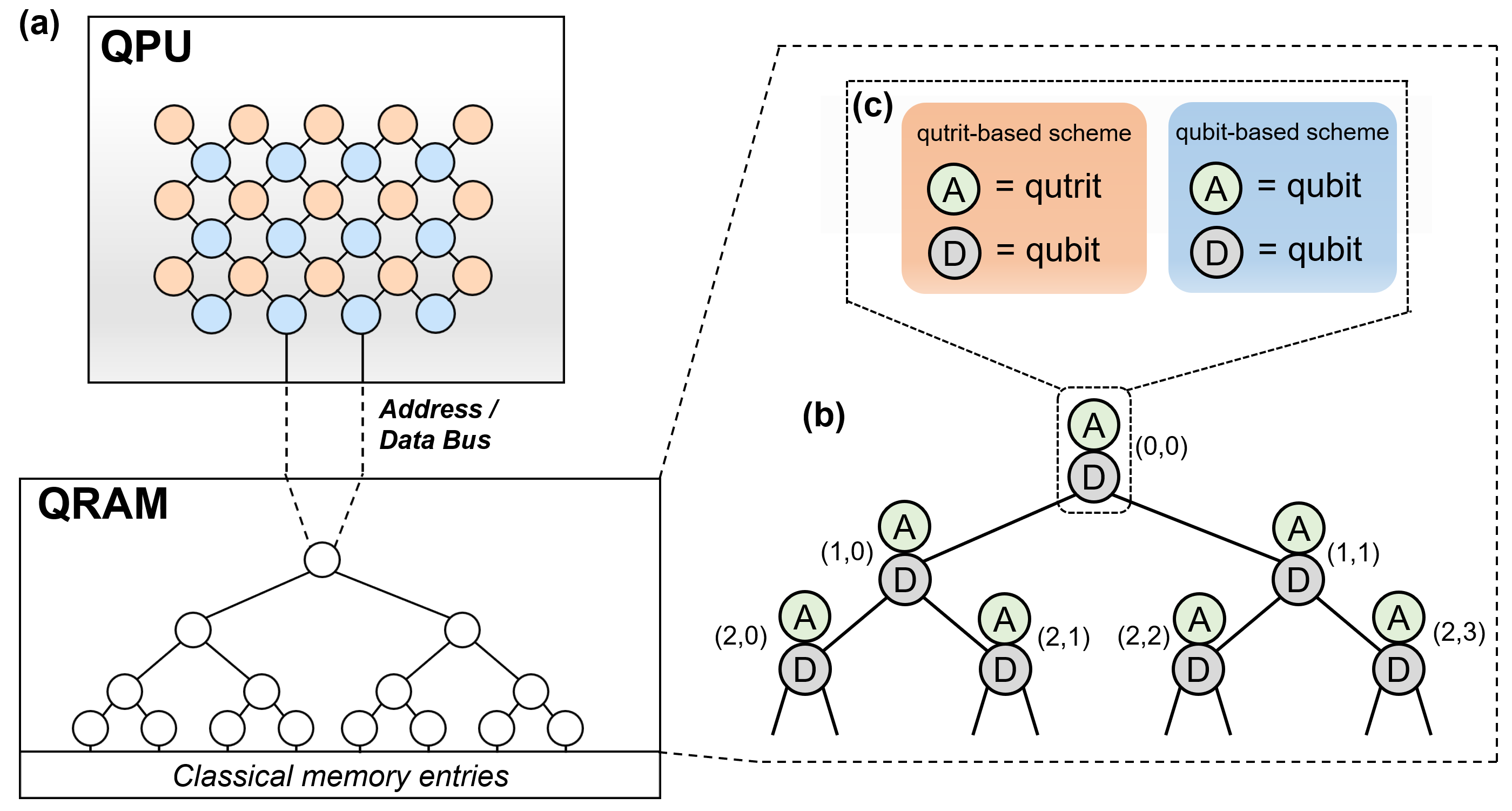}
    \caption{ (a) Illustration of a possible quantum computer architecture that includes a quantum processing unit (QPU) and a quantum random access memory (QRAM). The QPU connects the QRAM with a bus that transfers the address and the data state to the root of the QRAM. With a binary tree, QRAM connects with a classical memory. (b) The structure of the bucket-brigade QRAM. The structure is a full binary tree of $n$ layers. Each tree node is marked by a pair of integers, where the first denotes the number of the layer, the second the position within the layer. (c) The tree node has two cases in our protocol: a qutrit-based scheme and a qubit-based scheme. The qutrit-based scheme consists of an address qutrit and a data qubit. In the qubit-based scheme, the address qutrit is replaced by a qubit. The bottom of the tree is the classical memory. Each of the classical memory contains $k$ bits.}
    \label{fig:binary_tree}
\end{figure*}

While practical applications of QRAM are desirable, several challenges remain. The first challenge is the inherent difficulty in implementing the bucket-brigade QRAM architecture with fault tolerance. It requires an exponential number of qubits and non-Clifford gates \cite{ErrorResilience}, which cannot be transversally encoded in any quantum error correction codes \cite{newman2017limitations, zeng2011transversality, eastin2009restrictions}. For instance, a query from a surface-code-based QRAM would require an exponential number of T gates, making the required number of magic states intractable \cite{fowler2012surface}. A promising approach to achieve a practical application of the QRAM is to build it with high-quality physical qubits and error suppression methods such as error filtration \cite{errorfiltration} to further reduce errors. As a result, QRAM must operate with a limited number of qubits and cope with noisy qubits without any error correction.

Another considerable challenge lies in the issue of word length. Existing research has concentrated on scenarios where each data entry comprises a single bit, and an exact $n$-level QRAM is employed to independently route each of the $2^n$ bits. However, in practical applications, data entries usually consist of multiple bits, commonly known as the "word length" in the realm of classical computing. Denoting the word length as $k$, we need to load $2^nk$ bits in total. One could augment the number of QRAM levels to $n+\log k$ to individually address each bit, or extend the bandwidth or carry out sequential queries to handle this generalized data loading task, as referenced in \cite{ErrorResilience, QWQRAM2}. Nonetheless, these approaches carry significant drawbacks, including the requirement to construct at least $k$-fold qubits and/or confront a $k$-fold error rate. These challenges substantially impede the practicality of QRAM.

In this study, we aim to enhance the performance of generalized data-loading tasks, even when QRAM has limited resources. Firstly, we tackle the issue of word length by introducing a QRAM protocol which we refer to as the "parallel protocol." Our protocol successfully attains a time complexity of $O(n+k)$ for querying $2^n$ $k$-bit data items using an $n$-level QRAM, all without the necessity for additional bandwidth or an increase in the number of levels. This represents a substantial improvement in time complexity over previous works, specifically from $O(nk)$ to $O(n+k)$. Secondly, we extend this protocol to cater to any input size while utilizing a fixed-size QRAM. In scenarios where $2^{m+n}$ $k$-bit data items are queried with an $n$-level QRAM, the time complexity is reduced to $O(n+2^mk)$, representing another notable advancement over existing methods.

Our protocol not only surpasses existing methods in time complexity but also exhibits superior error scaling for the aforementioned tasks, leading to an overall enhancement in QRAM data access. To facilitate comparison between different protocols, we introduce the concept of a ``cost factor'' for the QRAM, offering a general measure of performance. Note that through the use of an error filtration method \cite{errorfiltration}, a trade-off can be established between the query rate and query error. Therefore, the cost factor serves as a measure of QRAM protocol quality, which is independent from the application of error filtration.

Finally, to demonstrate the benefits of the parallel protocol in error scaling, we conducted numerical simulations. The results intuitively demonstrate a slower increase in the error rate concerning the address length and word length, in comparison with the existing protocol. Although it might be contended that the QRAM is not likely to be infinitely scalable, our protocol significantly mitigates the prerequisites for constructing a practical QRAM. Given the indispensable role of QRAM in managing substantial volumes of data, our research could represent a promising step towards the implementation of practical QRAM in mid-term quantum computing systems, such as early fault-tolerant quantum computers.

\begin{figure*}[ht]
    \centering
    \includegraphics[width=1.0\linewidth]{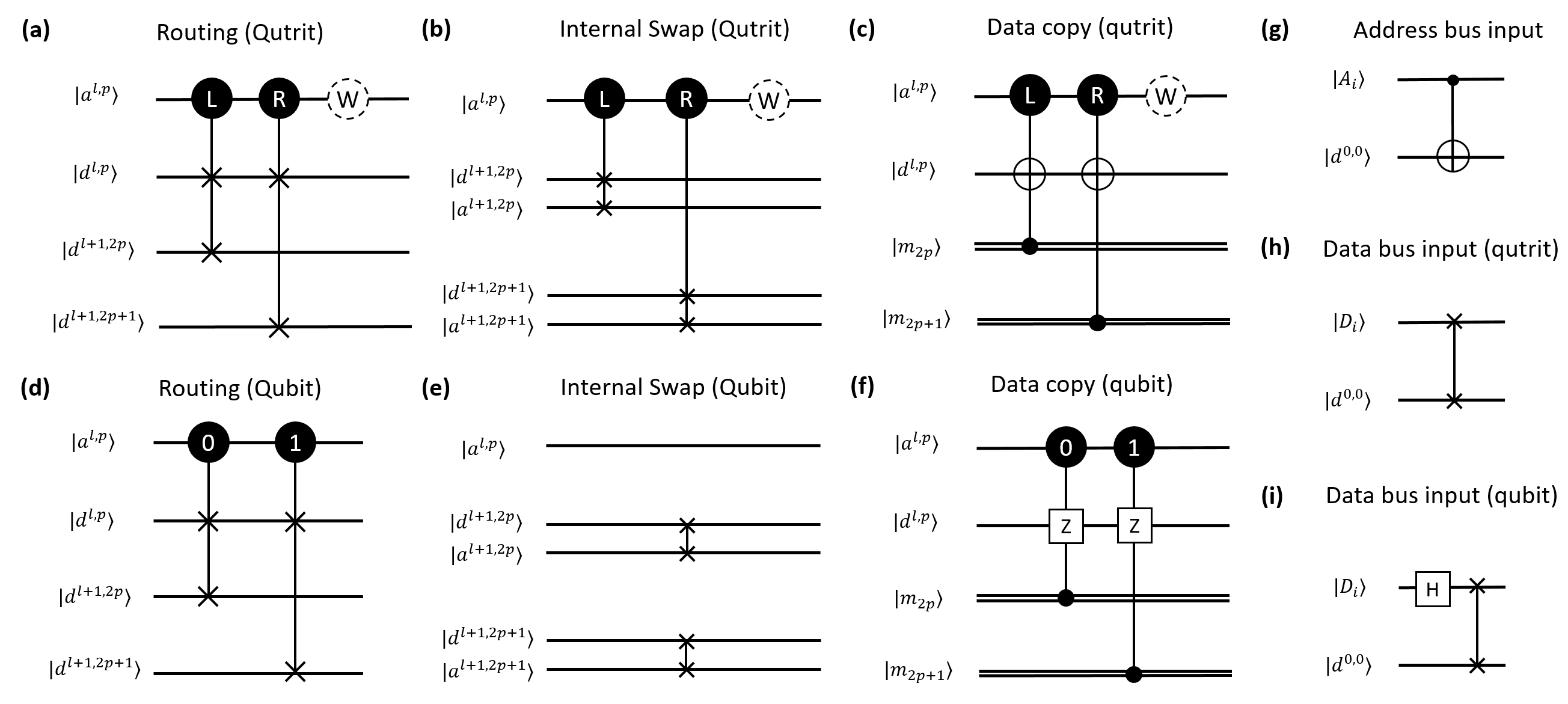}
    \caption{The fundamental operations in the QRAM, illustrated by quantum circuit model. (a) Routing operation in the qutrit scheme. (b) Internal swap in the qutrit scheme. Note that this operation is also controlled by the activeness of the parent node. (c) Data copy, denoted by $\text{M}_i$ in the qutrit model. (d)-(f) Same as (a)-(c), but in the qubit-based scheme. (g) The address bus input operation, which copies the address bus into the tree root. Note that this operation is identical in both schemes. (h) The data bus input operation in the qutrit-based scheme, which moves the data bus into the tree root. (i) Same as (h), but in the qubit-based scheme.}
    \label{fig:circuit}
\end{figure*}

\section{Background}

\subsection{Quantum random access memory}
We begin by introducing our definition of the data-loading task with an arbitrary word length. 

\begin{definition}[QRAM with an arbitrary word length]
 An $(n,k)$-QRAM acts on two quantum registers $|i\rangle_A$ and $|d\rangle_D$ , where register $A$ has $n$ qubits and $D$ has $k$ qubits, stores $2^n$ entries of $k$-bit classical data $m_i$ ($i\in[0,2^n-1]$), and implements $U_{\mathrm{QRAM}}$ such that 
\begin{equation}\label{eqn:U_QRAM}
U_{\mathrm{QRAM}}|i\rangle_A|d\rangle_D = |i\rangle_A|d\oplus m_i\rangle_D.
\end{equation}
\end{definition}

The $(n,k)$-QRAM determines the address length $n$ and word length $k$, which correspond to the size of the address and data register, with each classical data entry also being a $k$-bit string. Note that this study focuses on the bucket-brigade QRAM architecture by default. Other alternatives, such as the fanout architecture or QROM, have an error scaling that grows exponentially with the input size, especially with the address size $n$ \cite{ErrorResilience}. Therefore, considering that the QRAM is sensitive to errors, we do not consider them suitable candidates for a scalable QRAM architecture. Additionally, we have noticed a quantum-walk-based QRAM architecture \cite{QWQRAM, QWQRAM2}, which will be further discussed in Supplementary Information.

Fig.~\ref{fig:binary_tree} depicts a possible design of a quantum computer, encompassing a quantum processing unit (QPU) and a bucket-brigade QRAM. They are connected via an address bus and a data bus, which correspond to the QRAM's input registers $A$ and $D$ for the unitary $U_{\rm QRAM}$, respectively. The bucket-brigade QRAM is structured by a binary tree, of which tree nodes are composed of qudits (including qubit, qutrit, or more levels). A conceptual illustration of the QRAM structure is shown in Fig.~\ref{fig:binary_tree}(b). Hereinafter, we will mark each node by a pair of integers $(l,p)$, where $l$ denotes layer number and $p$ the position in the layer where $0\leq p\leq 2^l-1$. For example, the root node is $(0,0)$, and its left and right children are denoted as $(1,0)$ and $(1,1)$, respectively. 

Each node in the binary tree contains two qudits: an address qudit $A$ and a data qudit $D$. Generally, the address qudit marks the address route, while the data qudit transfers bits along the route. We classify the architecture into two schemes based on the type of qudit used for the address register: qutrit-based or qubit-based. As illustrated in Fig.~\ref{fig:binary_tree}(c), the qutrit-based scheme uses a qutrit for $A$ and a qubit for $D$, while the qubit-based scheme employs qubits for both $A$ and $D$. We propose this classification because the two schemes have different error scaling \cite{ErrorResilience}. The number of levels for the data qudit determines the number of bits that can be transferred at a time, namely the bandwidth of the QRAM. For simplicity, the data qudit is set to be one qubit (bandwidth one), and the higher-bandwidth case will be discussed later on.

The QRAM query process involves three phases: address setting, data fetch, and uncomputing. During the address setting phase, a path from the root of the tree to the bottom is established, corresponding to the address of the data entry being queried. In the data fetch phase, the data is first transferred from the data bus register to the tree root, then to the memory nodes for data bit retrieval, and finally returned to the bus. In the uncomputing phase, the address setting phase is reversed to return the system to its initial state. Note that the QRAM should be implemented unitarily, which is often considered a default requirement in many algorithms, such as those that involve reflection operations or uncomputation \cite{GrandUnification,QLSS}. This requirement allows the QRAM to not only map $|i\rangle|0\rangle$ to $|i\rangle|m_i\rangle$, but also to perform the inversion (See Supplementary Information for details). Therefore, it is important to completely uncompute all side effects that occur during the query process to ensure that the state of the QRAM is fully restored, with changes only to the state of the buses. 

\subsection{Fundamental operations in the QRAM process}
This section introduces the fundamental operations involved in QRAM queries in the quantum circuit model, which provides a clear way to model the time sequence and error. Fig.~\ref{fig:circuit} depicts the circuit for all operations, which include both versions for qutrit- and qubit-based schemes, respectively. The three levels of a qutrit are hereby denoted by $|L\rangle$, $|R\rangle$, and $|W\rangle$, while the states of a qubit are denoted by $|0\rangle$ and $|1\rangle$. The state of $A$ in node $(l,p)$ is denoted by $|a^{l,p}\rangle$, and the state of $D$ is denoted by $|d^{l,p}\rangle$. The state of the $i$-th qudit in the address and data bus is represented by $|A_i\rangle$ and $|D_i\rangle$, respectively. Lastly, $m_n$ denotes the $n$-th classical memory bit.

The \textit{routing} operation swaps the data qubit of a node with one of its children according to the state of the address qudit, which is illustrated in Fig.~\ref{fig:circuit}(a) and (d).

The \textit{internal swap} operation exchanges the data within a node. In the qubit-based scheme, the internal swap is simply a swap operation on the address qubit and data qubit, which is illustrated in Fig.~\ref{fig:circuit}(e). But in the qutrit-based scheme, it is conditioned by the activeness of the parent node. Fig.~\ref{fig:circuit}(b) is an example. Note that the internal swap is only necessary when the connections between qudits are limited, and there only allow routing operations between data qubits. If the connection is more flexible, one can directly apply the routing operation on the children nodes' address qudits instead of data qubits. Note that the internal swap operations are filled in the interval between other operations, and whether the connection has such a limitation will not affect the time sequence.

The \textit{data copy} operation retrieves the classical data to the data qubit by a classical-controlled operation, shown in Fig.~\ref{fig:circuit}(c) and (f). In the qutrit-based scheme, this is a classical-controlled controlled-not operation. In the qubit-based scheme, this is a controlled-Z operation where the data is encoded by the phase of the data qubit.

The \textit{address bus input} is identical in both schemes. It copies the $i$-th digit of the address bus to the root node using a CNOT operation. This input circuit will also be applied when in the uncomputing phase.

The \textit{data bus input} for the qutrit-based scheme is to move the $i$-th digit from the data bus to the root node using a swap operation. Not copying the data bus but moving it allows the uncomputing of the QRAM operation in the quantum algorithms. In the qubit-based scheme, we should apply a Hadamard gate on the data bus first and encode the bit on the phase of the qubit.

Routing, internal swap, and data copy operations can form into layered operations, which means that we will simultaneously apply this operation to an entire layer. For example, a layered routing $R^l_{t}$ is applied on all nodes in layer $l$. This parallelism is fundamental for the logarithm time scaling of the QRAM. Detailed descriptions of the fundamental operations and layered operations are shown in the Supplementary Information.

\begin{figure}[h]
    \centering
    \includegraphics[width=\linewidth]{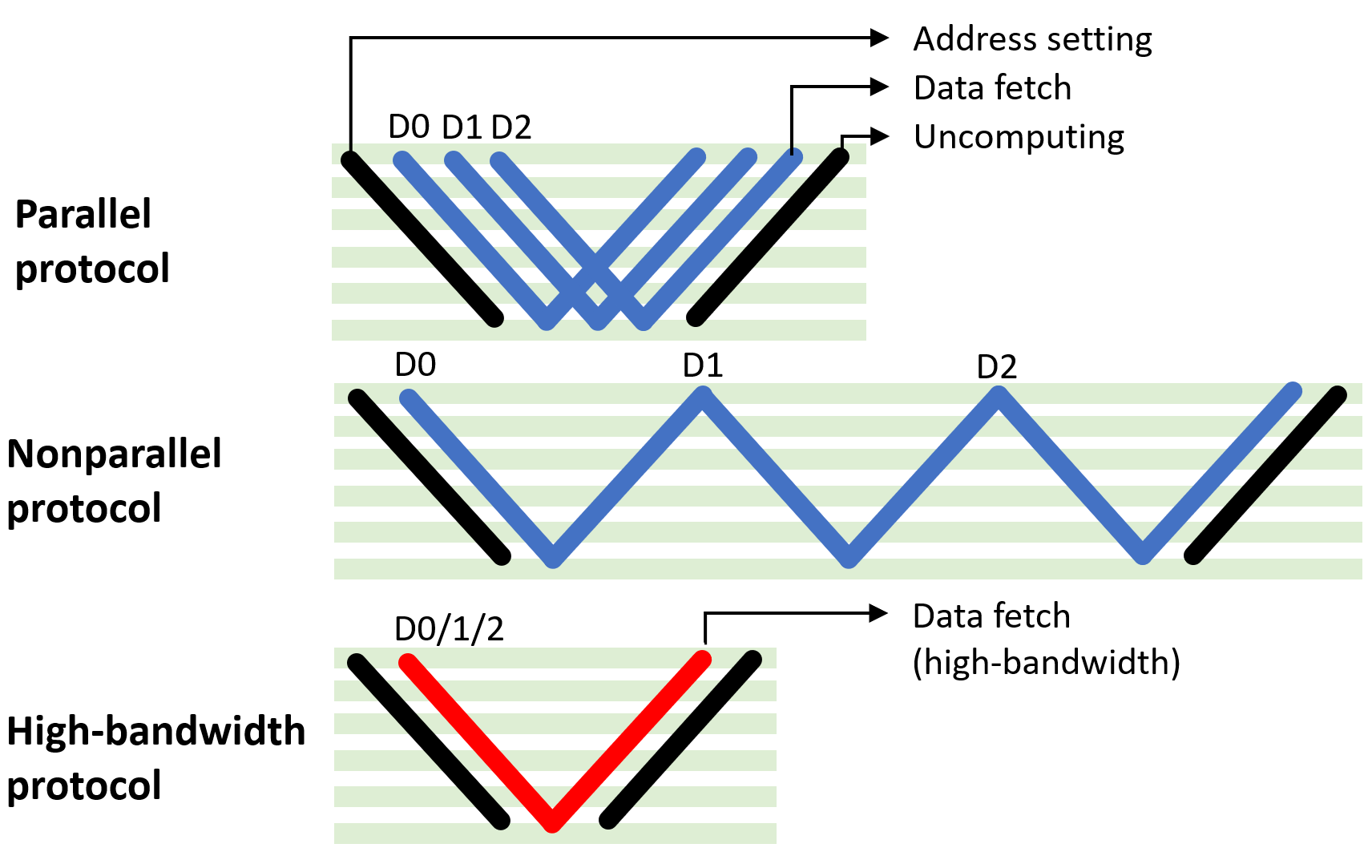}
    \caption{Schematic diagram illustrating the execution processes of different protocols. Horizontal bands represent the QRAM layers, and the slant from left to right represents the time sequence of data movement. The black slant represents the address setting phase or the uncomputing phase. The blue slant represents the data fetch phase of bandwidth one. The red slant is the high-bandwidth data fetch phase, wherein more than one digit of data can be transferred in a batch.}
    \label{fig:various_protocol}
\end{figure}

\section{The parallel protocol of QRAM}

\begin{figure*}[ht] 
    \centering
    \includegraphics[width=\textwidth]{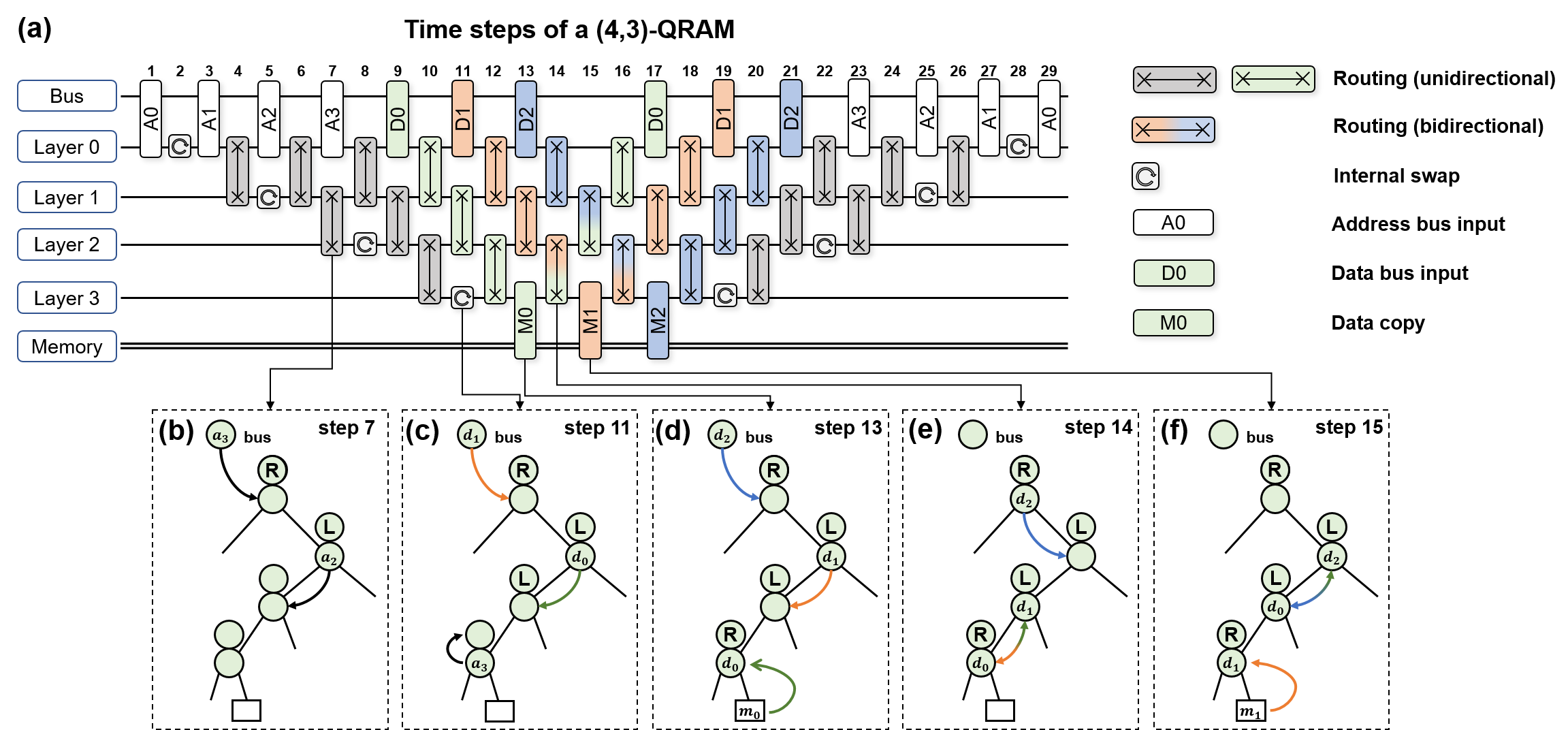}
    \caption{An example for the quantum circuit representation of a $(4,3)$-QRAM. (a) The time steps of a (4,3)-QRAM. Each horizontal line corresponds to a layer of the structure including the bus, QRAM, and memory. The legend of icons is shown on the right. Each icon can be a layered operation if the corresponding layers have more than one target. We mark the route of the three data bits with green, red, and blue colors. When two data bits are swapped in the neighbor layer (two routes collide), we use the gradient color to mark the bidirectional routing. (b)-(f) The behaviors of the data movement in five typical time steps of the branch with address 1001. Each circle represents a qudit, and each tree node consists of two qudits. The upper circle is the address qubit/qutrit, and the lower is the data qubit. The arrow denotes the movement of a bit.}
    \label{fig:fullcircuit}
\end{figure*}

\begin{table*}[ht]
\caption{Performance of different QRAM protocols.}
\label{tab:performances}
\begin{threeparttable}
\begin{tabular}{lccccc}
\toprule
Protocol Name  & Qubit number & \makecell[cc]{Time\\complexity} & \makecell[cc]{Error scaling \\(qutrit-based)}  & \makecell[cc]{Error scaling \\(qubit-based)} & \makecell[cc]{Cost factor (qutrit-based)\\ $T\epsilon_t/n\epsilon$} \\
\midrule
Nonparallel & $O\left(2^n\right)$ & $O\left(nk\right)$ & $O\left(kn^2\epsilon\right)$ & $O\left(kn^3\epsilon\right)$ & $k^2n^2$ \\
\makecell[ll]{High-bandwidth \\(bandwidth$=c$)}& $O\left(c2^n\right)$ & $O\left(nk/c\right)$ & $O\left(kn^2\epsilon\right)$ & $O\left(kn^3\epsilon\right)$ & $k^2n^2/c$ \\
Quantum-walk based \cite{QWQRAM2} & $O\left(n+k\right)$ & $O\left(n\log (n + k)\right)$ & - & - & - \\
Parallel (this paper) & $O\left(2^n\right)$ & $O\left(n+k\right)$ & $O\left(\left(n+k\right)n\epsilon\right)$ & $O\left(\left(n+k\right)n^2\epsilon\right)$ & $(n+k)^2$\\
\makecell[ll]{High-bandwidth parallel\\(this paper)}& $O\left(c2^n\right)$ & $O\left(n+k/c\right)$ & $O\left(\left(cn^2+kn\right)\epsilon\right)$ & $O\left(\left(cn^3+kn^2\right)\epsilon\right)$ \  & $cn^2+k^2/c+2kn$ \\
\bottomrule
\end{tabular}
\end{threeparttable}
\end{table*}

\subsection{Introducing parallelism in the QRAM process}
In this section, we introduce our QRAM protocol that implements an $(n,k)$-QRAM. We call this protocol the \textit{parallel protocol} because the main optimization comes from the parallelism achieved in the three phases of the QRAM process. 

This parallelism arises because we do not need to process each bit one by one. Instead, we can begin to process each address bit or data bit before its predecessor reaches its target. For instance, before an address qubit reaches its corresponding layer, another qubit can already enter the tree because the former route has already been carved. Moreover, different data qubits will not affect each other, even if their paths collide as one is entering and another is exiting the tree. With this principle, we are able to construct an operation sequence in which we only have to apply the address bus input and then data bus input operations compactly to push every bus digit into the tree. In Fig.~\ref{fig:various_protocol}, we illustrate the conceptual diagram for the execution processes of the parallel protocols. The horizontal bands represent the QRAM layers, and the slant lines represent the movement of the data. 

As a comparison, there are two trivial ideas for extending $(n,1)$-QRAM to $(n,k)$-QRAM. One is using the nonparallel version where each bit is queried sequentially, where we name it \textit{nonparallel protocol}. Another is extending the data qubit to multiple qubits (e.g. $m$-qubit) in each tree node, namely extending the bandwidth, where we name it \textit{high-bandwidth protocol}. These two protocols are also shown in Fig.~\ref{fig:various_protocol}.

Here we show an informal description of each phase of the parallel protocol. (For a formal description, see Supplementary Information.) The first is the address setting phase, where each digit of the address is pushed sequentially to a different level of the binary tree. Parallelism is achieved as each digit follows the path carved by its predecessors, and the next digits can be pushed in before the current digit reaches its destination. 

In the data fetch phase, each digit of the data bus travels from the root to the leaves, performs the classical controlled operation, and then travels inversely. Similar to the address setting phase, the digit can enter the tree after the last digit has moved to the next layer, which reduces the number of time steps required. Each data bit follows a zig-zag route, first moving downwards and then upwards, forming a folded trajectory that may intersect with the path of another data bit. When two routes collide, one routing operation is bidirectional, which allows for both upwards and downwards movement at the same time. As a result, after the address setting phase is completed, one can retrieve data from the memory through the same path without having to repeatedly set the address. When the word length is $k$, one only needs to repeat the data fetch $k$ times.

As an example, the quantum circuit representation of a (4,3)-QRAM is shown in Fig.~\ref{fig:fullcircuit}(a). In this representation, each wire represents all nodes in its layer, and the quantum operations (including routing, internal swap, and data copy) are layered operations that are executed simultaneously on all nodes in their layers. To make the movement of each data digit clear, different colors are used to distinguish the route taken by each data bit.

The digit transition process is illustrated through five typical time steps from panel (b) to (f) in Fig.~\ref{fig:fullcircuit}. The address setting phase demonstrates its parallelism starting from step 7, where A3 enters the tree and A2 moves from layer 1 to layer 2 simultaneously. Similarly, in the data fetch phase, step 13 (shown in panel (d)) allows D0, D1, and D2 to move simultaneously. Steps 14 and 15 (shown in panel (e) and (f), respectively) demonstrate the parallelism of the data fetch phase. In time steps 14 and 15, the collision of two color bands illustrates how a single operation enables the simultaneous upward and downward movement of two data digits. Note that this circuit representation applies to both the qutrit-based and qubit-based schemes, with each module set to its respective version.

\subsection{Performance analysis}
\subsubsection{Time complexity and qubit number}
The parallel protocol proposed in this paper improves the performance in terms of the time complexity, qubit number, error scaling, the overall cost, which will be analyzed in the next few sections. First, we show a summary of the comparison of the performance of the different protocols in Table~\ref{tab:performances}.

The time complexity of the parallel protocol for the $(n,k)$-QRAM is given below.
\begin{proposition}[Time complexity of the parallel protocol] The time complexity of the $(n,k)$-QRAM is $O(n+k)$ in the qutrit-based scheme and qubit-based scheme where $n$ is the address size, $k$ the data size.
\end{proposition}

For the parallel protocol, adding one more data qubit will only increase the number of time steps by 2, as the last data qubit has just left the tree root when the new data qubit enters the tree. Therefore, the time complexity of our protocol is linear in both $n$ and $k$, resulting in an overall time complexity of $O(n+k)$.

For the nonparallel protocol, the address setting phase takes the same $2n$ time steps. The data fetch phase has $k$ sequential data qubit's process, where the number of the time step is $2nk$. The uncomputing phase is the same as the address setting phase. The total number of time steps is $2nk+4n$, and the asymptotic complexity is $O(nk)$. In the high-bandwidth protocol with bandwidth $c$, a data qudit is constructed by $c$ qubits. Therefore, in the data fetch phase, we can allow it to transfer $k$ data bits in $k/c$ batches and obtain the time complexity $O(nk/c)$.

The time complexity of the parallel protocol shows a substantial speedup over the nonparallel protocol. Note that this complexity is asymptotically optimal with $O(2^n)$ qubits, as the limited bandwidth contributes at least $O(k)$, and the time for addressing each data bit is at least $O(n)$.

The qubit number of the parallel protocol and the nonparallel protocol is independent of the data length $k$, and the space complexity (qubit number) is $O(2^n)$, which is proportional to the number of the tree nodes of $n$ layers. The high-bandwidth protocol, as we have mentioned earlier, has $c$ qubits for the data qudits, so the qubit number is $O(c2^n)$.

\subsubsection{Error scaling}

Despite the qubit number and the time complexity, the parallel protocol mainly optimizes the error scaling of the QRAM.

\begin{proposition}[Error scaling of the parallel protocol]\label{pro:error}
    The fidelity of the $(n,k)$-QRAM is $F_1$ in the qutrit-based scheme and $F_2$ in the qubit-based scheme. We have $F_1\geq1-O((n+k)n\epsilon)$ and $F_2 \geq 1-O((n+k)n^2\epsilon)$, where $\epsilon$ is the error rate of each qudit and is sufficiently small.
\end{proposition}
The proof of the error scaling mainly inherits the idea in \cite{ErrorResilience}. To calculate the error scaling, one has to determine the fraction of \textit{good branches}, which is $\Lambda=1-nT\epsilon$ where $T$ is the execution time and $\epsilon$ is the error rate for each qubit. Then we have the error scaling as $1 - (2\Lambda-1)^2 = O(nT\epsilon)$. The error scaling consists of two components: the error rate of each node throughout the entire process, which is $O(T\epsilon)$, and the average number of branches affected by a single node's error.

To extend the proof from $(n,1)$-QRAM to $(n,k)$-QRAM, we only need to change the execution time in the proof from $n$ to $n+k$ (for the parallel protocol) or $nk$ (for the nonparallel protocol). For the high-bandwidth protocol with bandwidth $c$, the error rate for a single node is replaced by $O(c\epsilon)$. A detailed description of this extension is provided in Supplementary Information.

The results show that the parallel protocol preserves the error resilience, where the word length $k$ appends an extra $O(kn\epsilon)$ error to the system, and it finally has $O((n+k)n\epsilon)$ and $O((n+k)n^2\epsilon)$ error scaling in the qutrit-based scheme and qubit-based scheme, respectively. In contrast, the nonparallel has $O(kn^2\epsilon)$ error in the qutrit-based scheme and $O(kn^3\epsilon)$ in qubit-based, which increases the amount of error by a factor of $k$. Similarly, the high-bandwidth protocol increases the error of each qubit from $\epsilon$ to $c\epsilon$. Our results demonstrate that our proposed protocol is highly efficient for building a QRAM with a generalized input size. For a typical case where $n=k=32$, the error rate is improved from $O(n^3)$ to $O(n^2)$ and resulting in an approximately 16-fold improvement.

\subsubsection{Cost factor}

The error filtration method, which can suppress error in a black box unitary, is promising for reducing QRAM errors, particularly in situations where fault-tolerance is hard to achieved~\cite{errorfiltration}. Through error filtration, we can apply $T$ noisy black box unitary with $\epsilon$ error to achieve the same unitary but with $\epsilon/T$ error. Given a quantum algorithm that requires a QRAM query with a specific maximum error, this method can amend a highly noisy QRAM to meet the requirement by reducing the query speed. Therefore, we can define the cost factor of the QRAM by $T\epsilon_t / n\epsilon$, where $\epsilon_t$ is the error scaling and $\epsilon$ the error of a single qubit. A smaller cost factor represents better performance for a given number of QRAM levels, $n$. Note that although we use the qutrit-based error scaling, the qubit-based error scaling can produce the same result when setting the cost factor to be $T\epsilon_t / n^2\epsilon$. 

The asymptotic cost factors for various protocols can be easily calculated using the time complexity and error scaling, with results presented in Table~\ref{tab:performances}. The parallel protocol exhibits a cost factor of $(n+k)^2$, which surpasses both the nonparallel protocol $k^2n^2$ and the high-bandwidth protocol $k^2n^2/c$ (where $c\leq k$ always holds). 

\section{Parallel protocol for Arbitrary-sized data loading task\label{sec:application}}

The parallel protocol can also be applied to the arbitrary-sized data-loading task. 
\begin{definition}[Arbitrary-sized data loading]
 A $(n,m,k)$-data loading task is to load $2^{m+n}$-sized of $k$-bit data with an $n$-level QRAM.
\end{definition}

In previous works, this task is accomplished by a hybrid architecture that combines QRAM and QROM. A QROM is a quantum circuit embedding data loader which uses $2^n$ circuit depth and $n$ qubit. When the address length is $m+n$, we sequentially set the higher digit from $0$ to $2^m-1$ with the QROM, and query the lower $n$ digit with the QRAM. As a result, the hybrid architecture uses $2^m(n+k)$ time to finish the $(n,m,k)$-dataset loading when using the parallel QRAM protocol. As the hybrid architecture consists of $2^m$ sequential and independent QRAM queries, the error scaling is thus $O(2^m(n+k)n^2\epsilon)$ (here we only concentrate on the qutrit-based scheme). Structurally, the hybrid architecture is like the nonparallel protocol which accesses each subset of data sequentially. Inspired by the parallel protocol, we can design a parallel version of the hybrid architecture, which is called the \textit{hybrid-parallel} protocol.

\begin{figure}
    \centering
    \includegraphics[width=\linewidth]{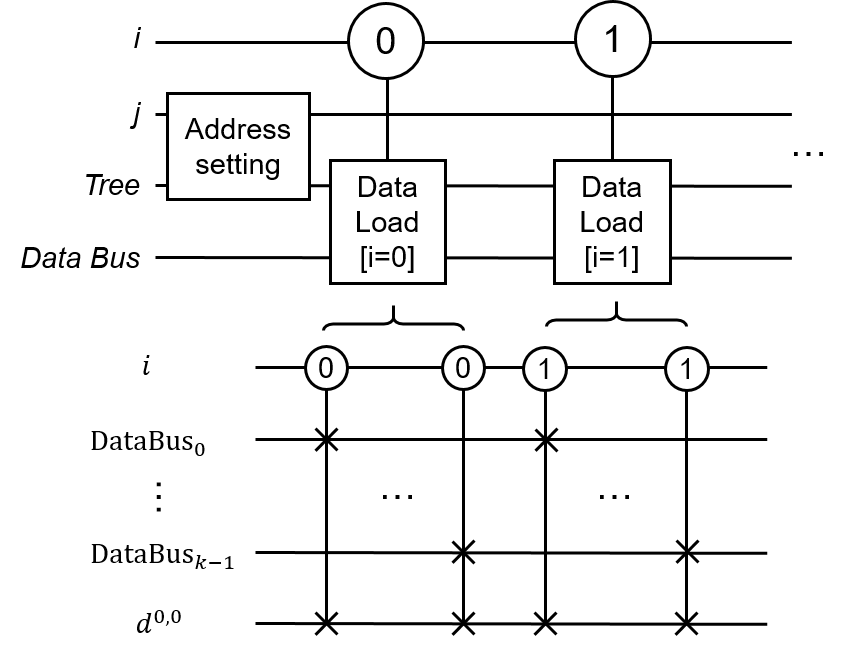}
    \caption{The quantum circuit for the hybrid-parallel protocol. The above circuit shows the control sequence, where we first perform the address setting phase according to the register $j$, then we perform the data fetch sequence with each $i$ as the controller. The detail for each data load module is shown at the bottom, where each data bus input is now controlled by register $i$. After completing all data from $i=0$, we then load from $i=1$ and restart from the first digit of the data bus.}
    \label{fig:extension}
\end{figure}

The quantum circuit is shown in Fig.~\ref{fig:extension}, where the $i,j$ are the address bus, \textit{Tree} is the QRAM system, and the \textit{Data Bus} is the data bus. Define $i$ and $j$ are the higher and lower digit of the address register, the task can be written as
\begin{equation}
    \sum_{i=0,j=0}^{2^m-1, 2^n-1}\alpha_{i,j}|i\rangle|j\rangle|d_{i,j}\rangle \rightarrow \sum_{i=0,j=0}^{2^m-1, 2^n-1}\alpha_{i,j}|i\rangle|j\rangle|d_{i,j}\oplus m_{i,j}\rangle.
\end{equation}

The first step is to perform the address setting phase, using $|j\rangle$ to initialize the QRAM tree. Then we modify the data fetch phase from a SWAP to a controlled SWAP so that the data bus input operation is controlled by the state of the higher digit. The first series load the $k$-bit when $i=0$, resulting in
\begin{equation}
\sum_{j=0}^{2^n-1}\alpha_{0,j}|0\rangle|j\rangle|d_{0,j}\rangle \rightarrow \sum_{j=0}^{2^n-1}\alpha_{0,j}|0\rangle|j\rangle|d_{0,j}\oplus m_{0,j}\rangle.
\end{equation}

And the next series load the $k$-bit when $i=1$, etc. After iterating over all $2^m$ higher digits, we perform the uncomputing to recover the QRAM system. 

The performance is as follows.
\begin{proposition}[Performance of the hybrid-parallel protocol on $(n,m,k)$-dataset loading] The time complexity of the hybrid-parallel protocol with $(n,k)$-QRAM on $(n,m,k)$-data loading task is $2^mk+n$. The error scaling is $O((2^mk+n)n\epsilon)\sim O(2^mnk)$.
\end{proposition}
The time for loading $2^m$ data is independent with $n$ so that the time complexity is thus $2^mk+n$. Similar to above, the error scaling is $O((2^mk+n)n\epsilon)\sim O(2^mnk)$. Compared with the hybrid architecture, this hybrid-parallel protocol significantly reduces the time and error. 

Moreover, we want to discuss the parallel protocol's ability to load sparse data. The hybrid-parallel protocol treats the data as $2^n$ number of dense binary strings where each is of $2^mk$-sized. This also can be applied to the case when the data is sparsely encoded, where we only have to adjust $2^mk$ to some $k'$ and treat the original data loading task into the $(n,k')$-QRAM task. For example, a hash table is an array of buckets where each bucket may contain at most $M$ elements. The parallel protocol can use an $n$-level QRAM and $O(n+Mk)$ time to access all data and error scaling $O((n^2+nMk)\epsilon)$.

\section{Extension: The high-bandwidth parallel protocol}
The parallel protocol can be extended to the \textit{high-bandwidth parallel protocol} by adding the number of data qubits in a node. For the high-bandwidth protocol, the query process is similar to the parallel protocol. In the data fetch phase, several bits are transferred in a batch. When the bandwidth is extended to $c$, the time complexity in the data fetch phase is $n + k/c$ for an $(n,k)$-QRAM instance. The address setting phase does not greatly change, because one still requires $O(n)$ time to transfer the address bits from the root to the leaves. Therefore, the total time complexity of the high-bandwidth parallel protocol has the time complexity $O(n+k/c)$.

The error scaling of this high-bandwidth extension can also be calculated similarly as above. The error in each node is increased to $c\epsilon$, therefore the error in the data fetch becomes $O((n+k/c)nc\epsilon) = O((cn^2+kn)\epsilon)$ in the qutrit-based scheme, and $O((n+k/c)n^2c\epsilon) = O((cn^3+kn^2)\epsilon)$ in the qubit-based scheme.

The high-bandwidth parallel protocol shows a trade-off between a faster query speed and a lower error rate. Additionally, we observe that when $n$ and $k$ are fixed, setting $c=k/n$ enables an optimal QRAM configuration without considering the number of qubits. This also implies that increasing the bandwidth without increasing the number of levels could also potentially lower the cost of the QRAM.

\section{Numerical simulation\label{sec:numerical}}

We study the fidelity of the parallel protocol with different address sizes and word lengths by a classical simulator and compare it with the nonparallel protocol. The simulation method mainly follows \cite{ErrorResilience}, where we extend it to simulate the QRAM with arbitrary word length.

The simulation starts from randomly generated QRAM memory and $d$ superposition bus input, that is
\begin{equation}
    \sum_i\alpha_i|a_i\rangle|z_i\rangle.
\end{equation}
During the execution of the QRAM circuit, we apply the damping and/or depolarizing error channels to each working qudit. After the process finishes, the bus will be entangled with the QRAM system. Then we partial-trace the QRAM system and obtain a mixed state in the bus, that is
\begin{equation}
    \sum_j p_j|\psi_j\rangle\langle \psi_j|.
\end{equation}
Finally, we calculate the fidelity $F$ of the output where
\begin{equation}
    F=\sum_{i,j} p_j|\alpha_i|^2 |\langle a_i, z_i\oplus m_{a_i}|\psi_j\rangle|^2.
\end{equation}
The above process is repeated $10^4$ times to obtain the average fidelity and the standard error.
\begin{figure}
\begin{overpic}[width=\linewidth]{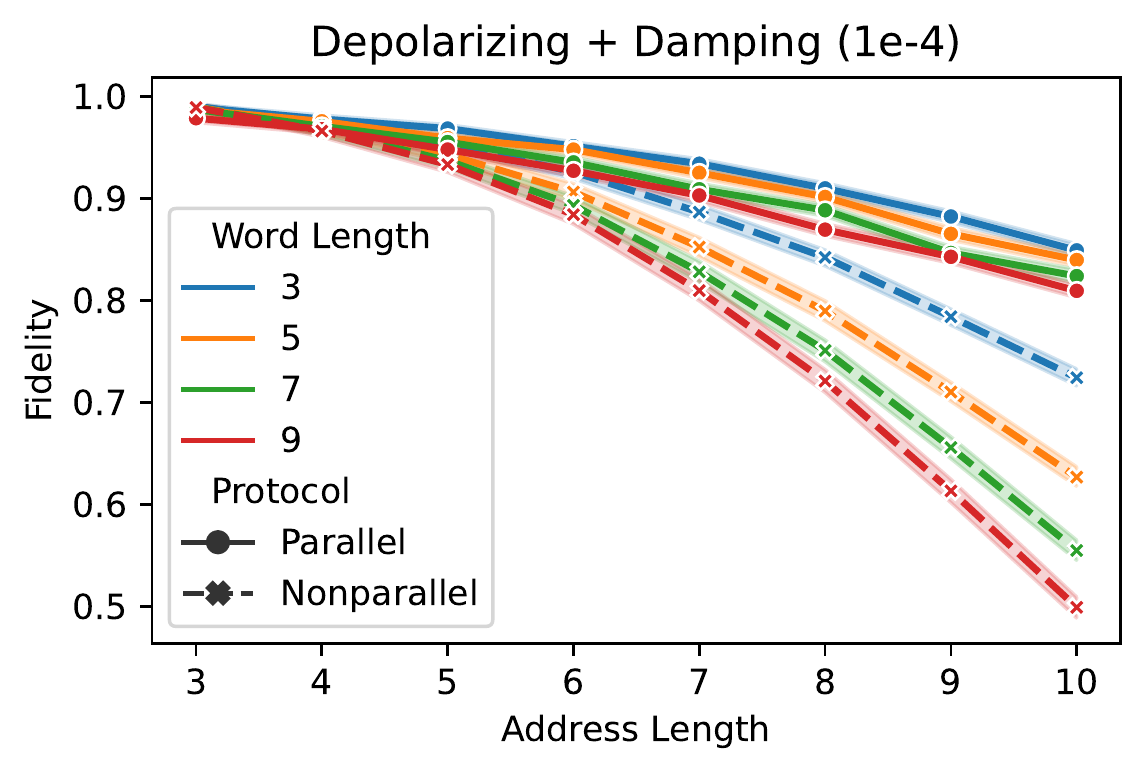}
  \put(1,62){\textbf{(a)}}
\end{overpic}
\begin{overpic}[width=\linewidth]{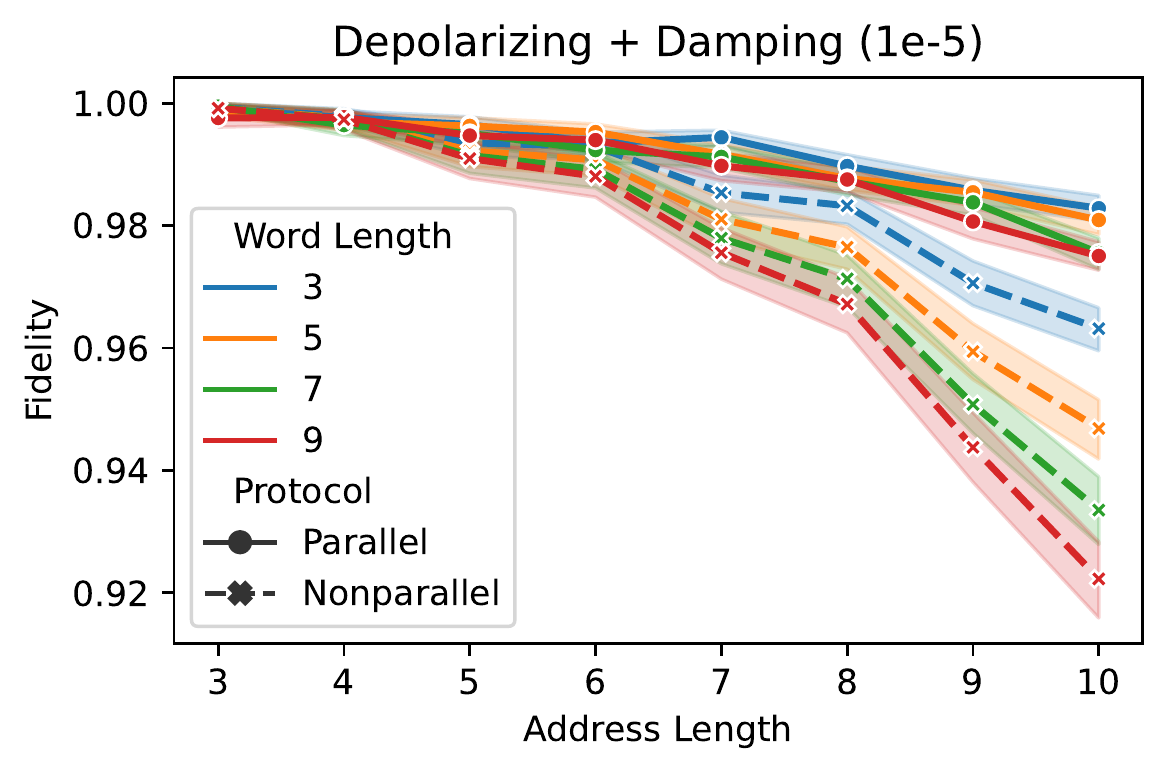}
  \put(1,62){\textbf{(b)}}
\end{overpic}
\caption{Results of the classical simulation of QRAM, showing the relation between the fidelity and the address length. (a) Each work qubit/qutrit has 1e-4 damping and depolarizing noise. Two methods are compared: parallelized (solid lines) and non-parallelized method (dotted lines) when retrieving data with different word lengths. Comparing the dotted line with the solid of the same word length, a substantial fidelity improvement of the parallel protocol can be revealed. (b) Same as (a), but with 1e-5 damping and depolarizing noise.}
    \label{fig:compare}
\end{figure}

We compare the parallel and nonparallel protocol for the same $(n,k)$-QRAM. As stated above, these two protocols have the same structure but different execution processes. In Fig.~\ref{fig:compare}, we plot the change of fidelity with different word lengths. The solid lines and the dotted lines with the same color represent the same error and size configuration with parallel and nonparallel protocol, respectively. Both figures with $10^{-4}$ and $10^{-5}$ error strength observe that the parallel protocol has substantial improvement in fidelity compared to the nonparallel protocol.


\begin{figure*}[ht]
\begin{overpic}[width=0.34\linewidth]{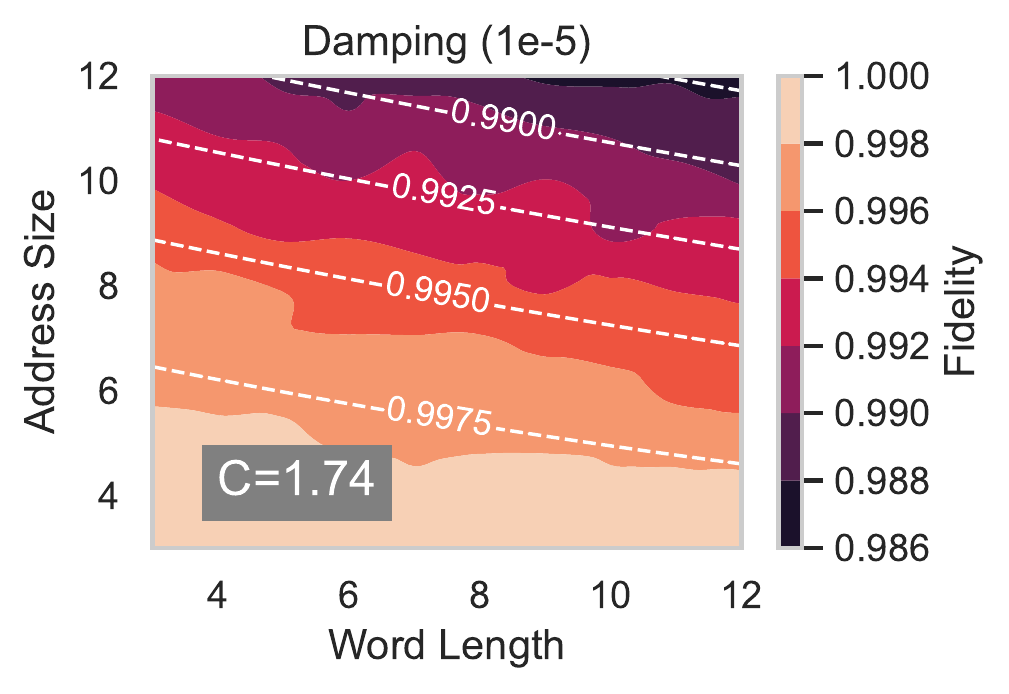}
  \put(1,70){\textbf{(a1)}}
\end{overpic}
\begin{overpic}[width=0.32\linewidth]{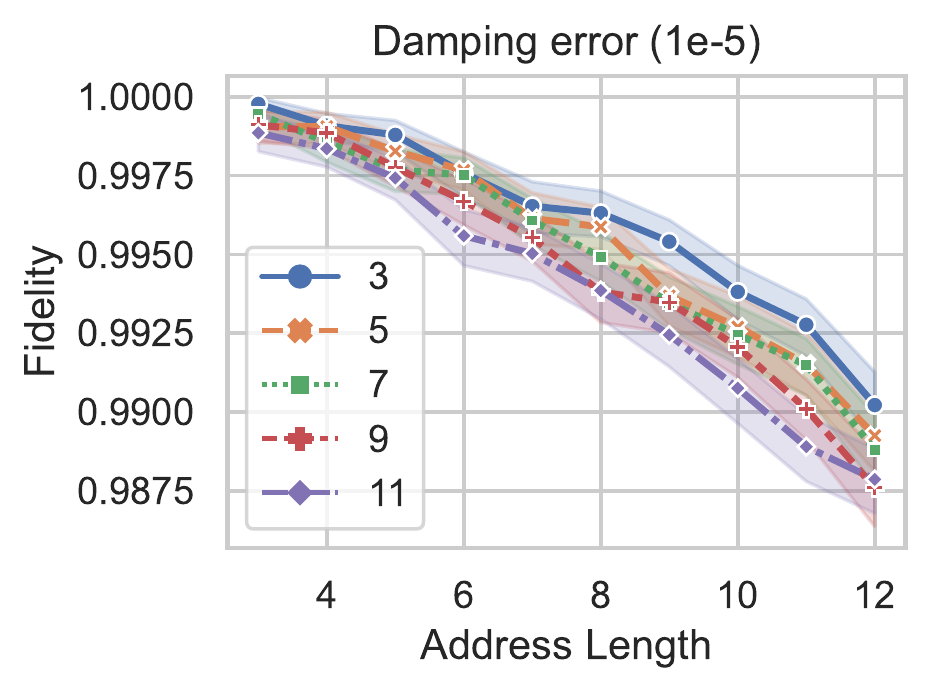}
  \put(1,74){\textbf{(a2)}}
\end{overpic}
\begin{overpic}[width=0.32\linewidth]{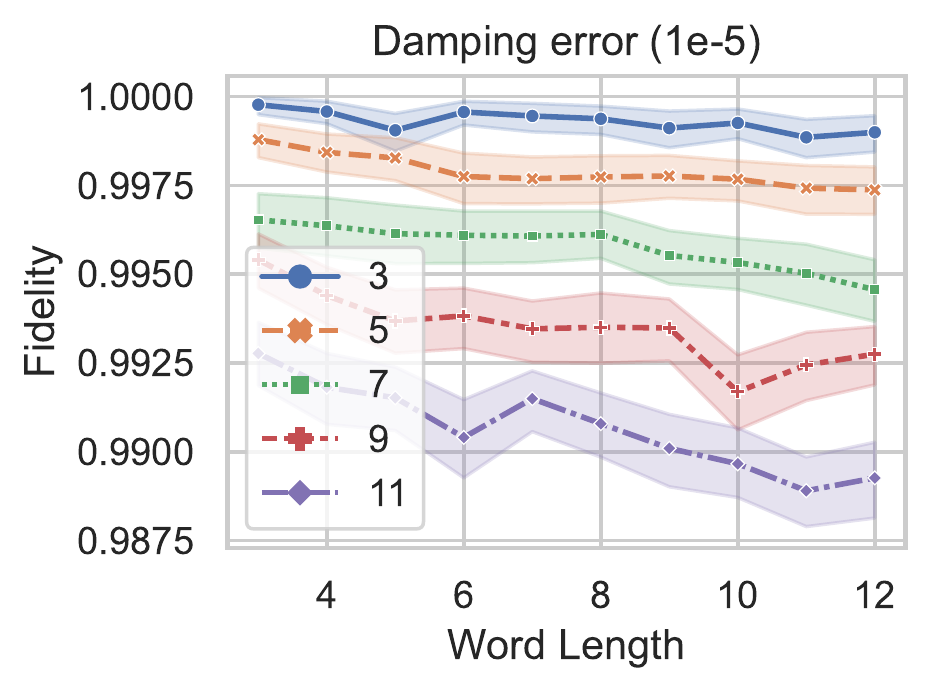}
  \put(1,74){\textbf{(a3)}}
\end{overpic}
\begin{overpic}[width=0.34\linewidth]{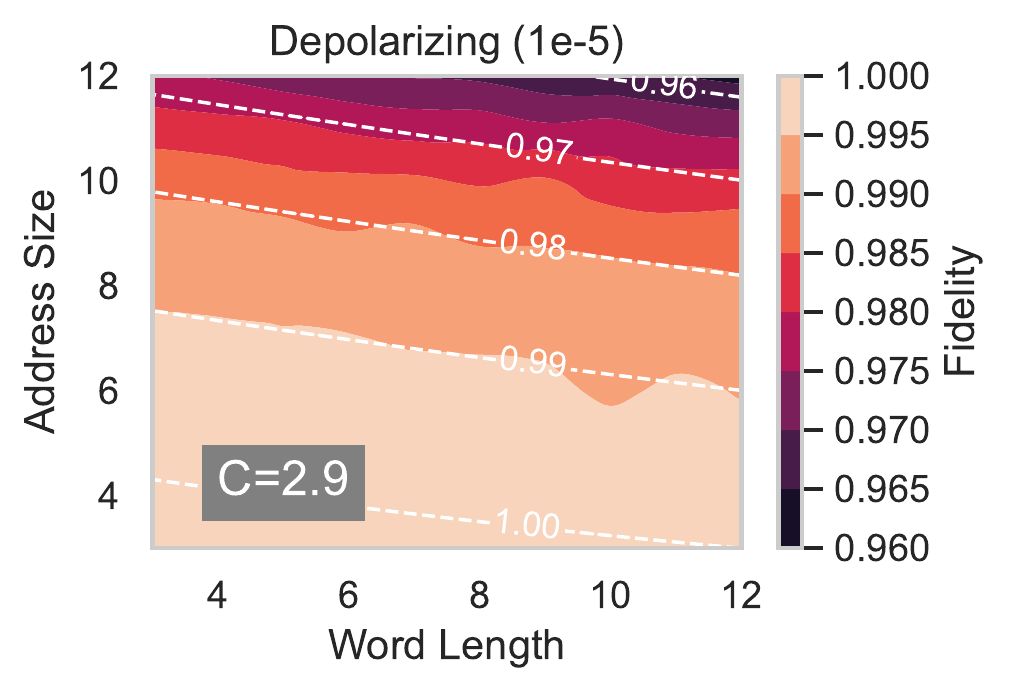}
  \put(1,70){\textbf{(b1)}}
\end{overpic}
\begin{overpic}[width=0.32\linewidth]{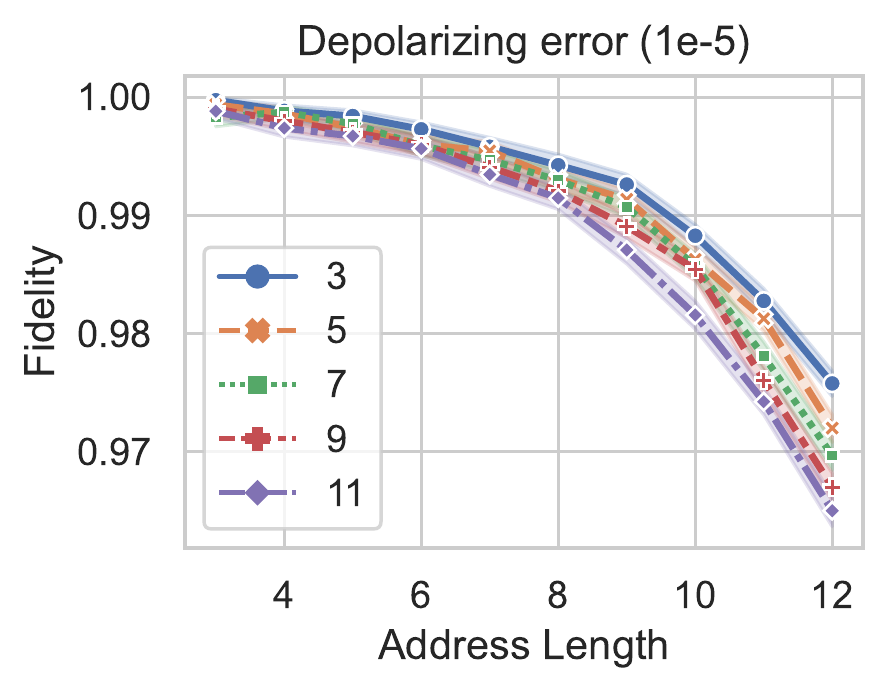}
  \put(1,74){\textbf{(b2)}}
\end{overpic}
\begin{overpic}[width=0.32\linewidth]{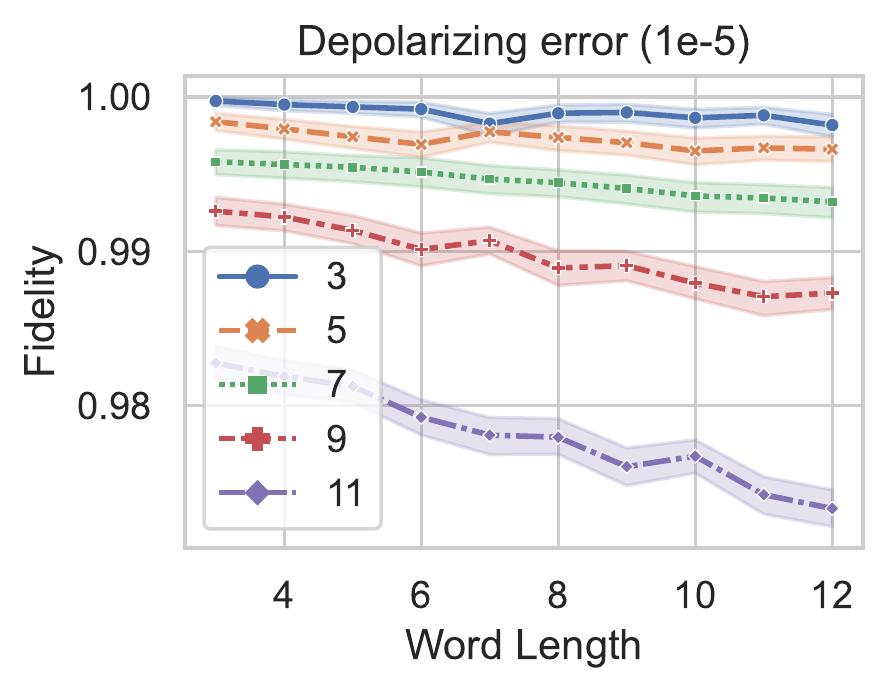}
  \put(1,74){\textbf{(b3)}}  
\end{overpic}  
\begin{overpic}[width=0.34\linewidth]{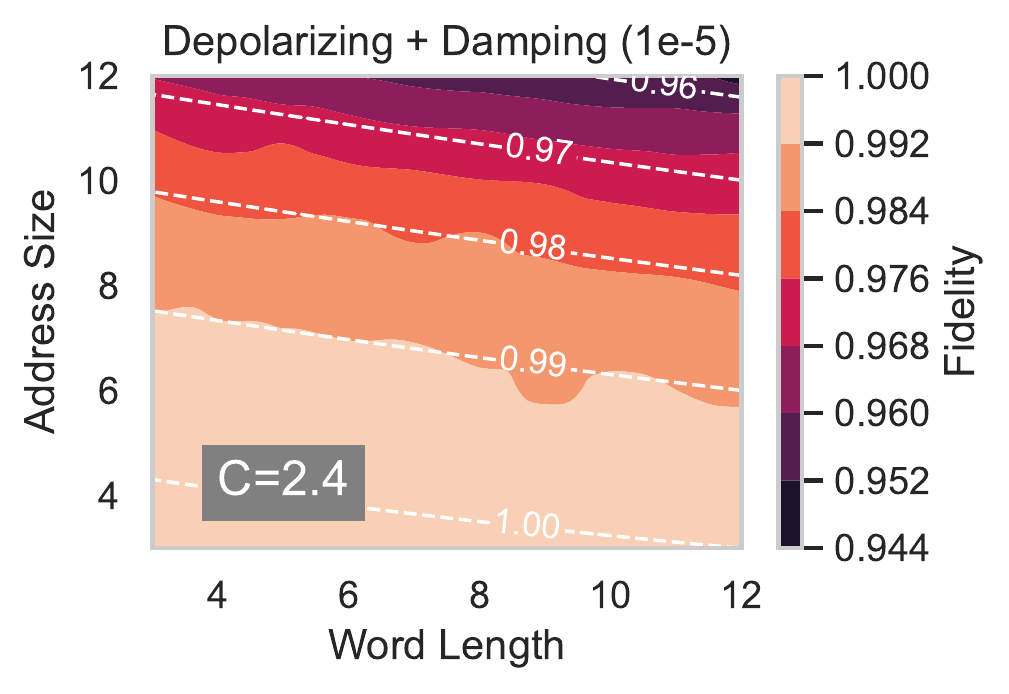}
  \put(1,70){\textbf{(c1)}}
\end{overpic}
\begin{overpic}[width=0.32\linewidth]{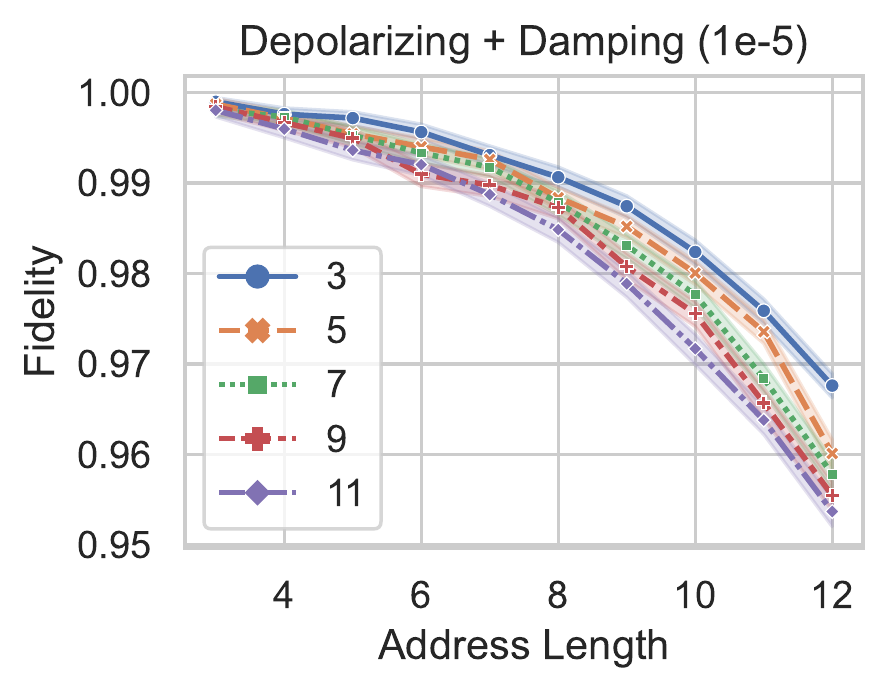}
  \put(1,74){\textbf{(c2)}}
\end{overpic}
\begin{overpic}[width=0.32\linewidth]{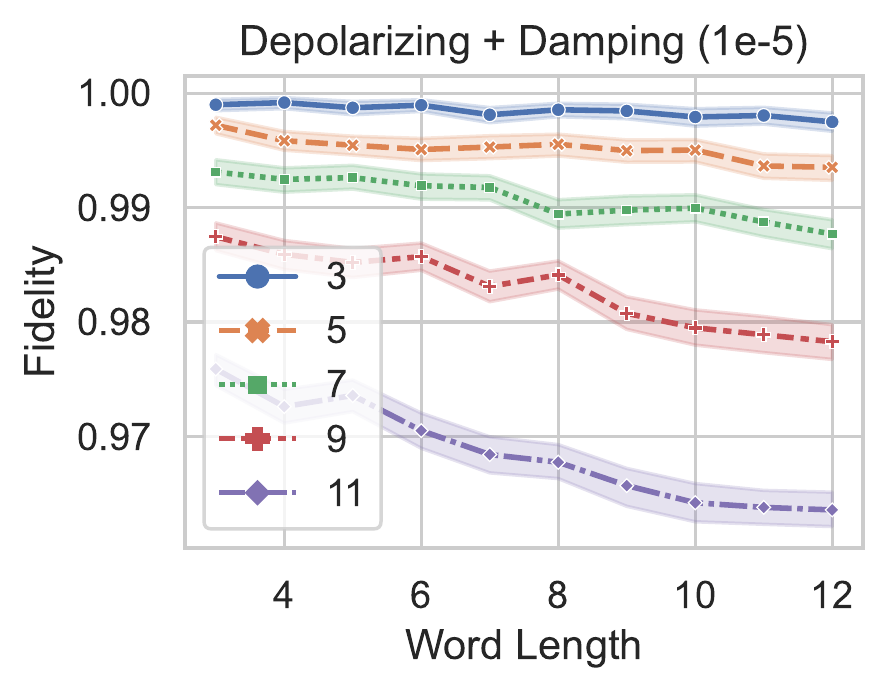}
  \put(1,74){\textbf{(c3)}}
\end{overpic}
\caption{Fidelity for various error channel types. (a1-a3) Only a damping channel is applied. (b1-b3) Only a depolarizing channel is applied. (c1-c3) Both a damping and a depolarizing channel are applied. (a1-c1) Heatmaps illustrating the relationship between fidelity, address size $n$, and word length $k$ for different error channels. The contour line depicts the relation between $n$ and $k$ based on the fitting formula. (a2-c2) Fidelity as a function of address size, with each line corresponding to a different word length as shown in the legend. (a3-c3) Fidelity as a function of word length, with each line corresponding to a different address size as shown in the legend.}
\label{fig:error}
\end{figure*}

Also, we test the change of QRAM fidelity with a certain error channel with the address size and word length, whose results are illustrated in Fig.~\ref{fig:error}. The error channel is the damping channel where $\gamma=10^{-5}$ (details about damping channel) in subfigure (a) series, the depolarizing channel where $p=10^{-5}$ in subfigure (b) series, and the compound of the above two channels in subfigure (c) series. Both the address size $n$ and word length $k$ range from 3 to 12, and the contour map is plotted in Fig.~\ref{fig:error}(a1), (b1) and (c1). The contour line is fitted from the following formula ($A$ and $C$ are fitting variables):
\begin{equation}
    F=1-A(Cn^2+nk)\epsilon.
\end{equation}
The variable $C$ represents the relative magnification of the variable affected by the two dominated terms $n^2$ and $nk$, where fitting results show the correct relation on $F$ with $n$ and $k$. In Fig.~\ref{fig:error}(a2), (b2) and (c2), different lines represent the corresponding word lengths; in Fig.~\ref{fig:error}(a3), (b3) and (c3), different lines represent the corresponding address sizes. These figures imply a mild and linear dependency of the fidelity on the word length, and the increasing $n$ will induce a quadratic increase in the error rate.

\section{Summary and Outlook\label{sec:summary}}

In summary, we propose an optimized protocol which is named parallel protocol. The parallel protocol implements $(n,k)$-QRAM with $O(2^n)$ qubits or qutrits, $O(n+k)$ execution time complexity, and $O((n+k)n\epsilon)$ error rate. Other two possible protocols, the nonparallel protocol, and the high-bandwidth protocol are compared with the parallel protocol, where the time complexity and the number of qubits and qutrits of the parallel protocol have great advantages over its counterparts. More importantly, the error scaling is substantially improved from $O(n^2k\epsilon)$ to $O((n+k)n\epsilon)$. For large $k$ such as $n\sim k$, the protocol effectively improves the error from $O(n^3)$ to $O(n^2)$. Numerical simulations are performed and demonstrate the improvement of the error performance of the parallel protocol.

The application to the arbitrary-sized data loading task is also discussed. When the number of data entries is $2^{m+n}$ where only $n$-level QRAM is given, multiple accesses of QRAM are required. We propose the hybrid-parallel protocol, loading the QRAM with $O(2^mk+n)$ time and $O((2^mk+n)n\epsilon)$ error scaling. This is also a significant improvement over the hybrid architecture that combines the QRAM and QROM.

The cost factor, a measure of the quality of various QRAM protocols is also proposed. By combining the noisy QRAM and the error filtration method, one can make a trade-off for the fidelity and the time complexity of a QRAM. In terms of the cost factor, the parallel protocol also shows a substantial improvement. We also discuss the high-bandwidth parallel protocol, which is a natural extension of the parallel protocol by extending the bandwidth and making use of the parallel retrieval of data. As a major variable, a higher bandwidth provides us with a possible way to reduce the cost of a QRAM by adding more qubits.

As the classical data loading task is essential for almost all quantum algorithms that require a large amount of real-world data, we should keep improving the performance of the QRAM. Even the best way for loading more data bits into the quantum computer is to build a QRAM that can retrieve every data bit in parallel, however, it is always difficult to build a QRAM as large and high-fidelity as we want. As a possible solution, our protocol can be applied to a finite-sized QRAM for big data with better performance, which suggests a promising application for QRAM in mid-term quantum computers.

\section*{Data Availability}
The data that support the plots within this paper are available from Z.-Y.~C. upon reasonable request.

\section*{Competing Interests}
The authors declare no competing financial or non-financial interests.


\begin{acknowledgments}
This work was supported by the National Natural Science Foundation of China (Grant No.~12034018), and Innovation Program for Quantum Science and Technology No.~2021ZD0302300.
\end{acknowledgments}

\bibliography{apssamp}

\clearpage
\setcounter{table}{0}
\renewcommand{\thetable}{S\arabic{table}}%
\setcounter{figure}{0}
\renewcommand{\thefigure}{S\arabic{figure}}%
\setcounter{section}{0}
\setcounter{equation}{0}
\renewcommand{\theequation}{S\arabic{equation}}%

\onecolumngrid

\begin{center}
{\large \bf Supplementary Information:
 Efficient and Error-Resilient Data Access Protocols for a Limited-Sized Quantum Random Access Memory }\\
\vspace{0.3cm}
\end{center}

\setcounter{page}{1}

\section{Theory of Quantum Random Access Memory}

The main structure of the quantum random access memory (QRAM) with the bucket brigade architecture has been briefly introduced in the main text. In this section, we will give a formal description of the QRAM, including the structure, the operation, and the query process.

\subsection{Definitions and structure}
First we define the QRAM tree. For an $n$-level QRAM, the tree is a $n$-layer full binary tree. Each tree node is a routing unit which has two qudits: one qudit to store the address and another to allow movement of the data. Generally we have two kinds of models. One is that we use a qutrit to mark the status of the address: ``left'', ``right'' and ``wait''. ``Left/Right'' means the direction of an opened route, and ``wait'' means that this node is not activated. Another model is that the address is only marked by a qubit: only left and right are encoded. Either model can be considered as a valid model of the bucket brigade QRAM which holds the error resilient property. The data qudit is free to select, where it only represents the ``bandwidth'', namely the number of bits that can be transferred through this node at a time. A typical case is to use one data qubit. The high-bandwidth case, where more data qubits are applied in one node, is easy to extend from the bandwidth 1 case, which will be discussed later.

To identify a qudit in the QRAM tree, a pair of integers $(l,p)$ are applied in the main text to uniquely locate the position of a qudit. Now we can use $|a^{(l,p)}\rangle$ or $|d^{(l,p)}\rangle$ to represent the state of any address or data qudit. 

In the qutrit-based scheme, the three states (left, right and wait) of the address qutrit are $|L\rangle$, $|R\rangle$ and $|W\rangle$, respectively. In the qubit-based scheme, the address qubit encodes $|L\rangle$ (left) by $|0\rangle$ and $R\rangle$ (right) by $|1\rangle$.

\subsection{Fundamental operations}
The fundamental operations involved in the QRAM query process are \textit{routing}, \textit{internal swap}, \textit{data copy}, \textit{address bus input} and \textit{data bus input}. Here we provide formal expressions of all these operations.

Most operations have two versions: the qutrit-based and the qubit-based schemes. To identify these two versions, we mark the operation with a subscript $t$ for ``trit'' and $b$ for ``bit''. For a default case (where no subscript is given), the operation is for the qutrit-based scheme.

\subsubsection{Routing}
The routing operation $R_{t/b}$ is performed on two qudits in one node ($|a^{(l,p)}\rangle_1$ and $|d^{(l,p)}\rangle_2$), and the data qubits in left/right children nodes $|d^{l+1, 2p}\rangle_3$ and $|d^{l+1, 2p+1}\rangle_4$. The position $2p$ and $2p+1$ can be directly obtained from the property of the full binary tree. The subscripts 1, 2, 3, and 4 are used to mark different qudits.

For the qutrit-based scheme, we have

\begin{equation}
R_t = |L\rangle\langle L|_1\otimes \textsc{Swap}^L_{2,3}\otimes I_{4}+|R\rangle\langle R|_1\otimes \textsc{Swap}^R_{2,4}\otimes I_{3}+|W\rangle\langle W|_{1}\otimes I_{2,3,4}.
\end{equation}
Here $\textsc{Swap}_L$ means the swap operation on $|d^{(l,p)}\rangle$ and $|d^{l+1, 2p}\rangle$; $\textsc{Swap}_R$ means the swap operation on $|d^{(l,p)}\rangle$ and $|d^{l+1, 2p + 1}\rangle$. $I_n$ represents the identity at the qudit $n$.

For the qubit-based scheme, we have
\begin{equation}
R_b = |0\rangle\langle 0|_1\otimes \textsc{Swap}^L_{2,3}\otimes I_{4}+|1\rangle\langle 1|_1\otimes \textsc{Swap}^R_{2,4}\otimes I_{3}.
\end{equation}

\subsubsection{Internal swap}
The internal swap operation $I_{t/b}$ is used to encode the address qudit by the data qubit. In the QRAM query process, the address qudits in each layer should be encoded by the state of the address bus, respectively. This state is passed from the address bus to the corresponding layer through the path of data qubits in the activated nodes. After this state is stopped at the data bus of the destination layer, it should be passed to the address qudit in the same node. The word ``internal'' means the state is transferred inside the node.

For the qutrit-based scheme, we have
\begin{equation}
I_t=|L\rangle\langle L|_1\otimes \textsc{intSwap}_{2,3}+|R\rangle\langle R|\otimes\textsc{intSwap}_{4,5}+|W\rangle\langle W|_1\otimes I_{2,3,4,5}.
\end{equation} 
This operation involves four qudits: the address qudit in a node $|a^{(l,p)}\rangle_1$ and the address and data qudits in both two children nodes ($|a^{(l+1,2p)}\rangle_{2}$, $|d^{(l+1,2p)}\rangle_{3}$, $|a^{(l+1,2p+1)}\rangle_{4}$, $|d^{(l+1,2p+1)}\rangle_{5}$. 
Note that $\textsc{intSwap}_{2,3}$ is a ``swap'' between a qubit and a qutrit, which implements the mapping
$$|W,0\rangle \mapsto |L,0\rangle,$$
and
$$|W,1\rangle \mapsto |R,0\rangle.$$
Also, it is worth noting that the internal swap in the qutrit-based scheme is conditioned by the activeness in the address node in the last layer.

For the qubit-based scheme, the internal swap operation is simple, which is 
\begin{equation}
I_b=\textsc{Swap}_{2,3} \otimes \textsc{Swap}_{4,5}.
\end{equation}
The indices of qudits remain the same as above.

A possible implementation for the qutrit-based internal swap is as follows. First, let the parent node excite the address qutrit of the corresponding child from $|W\rangle$ to $|L\rangle$, that is 
\begin{equation}
|L\rangle\langle L|\otimes X_L + |R\rangle\langle R|\otimes X_R + |W\rangle\langle W|\otimes  I.
\end{equation}
Here $X_{L/R}$ means an $X$ flip between $|W\rangle$ and $|L\rangle$ in the left/right address qutrit. This is a mapping of 
$$|W,0\rangle \mapsto |L,0\rangle,$$
and
$$|W,1\rangle \mapsto |L,1\rangle.$$
This process is like an ``activation'' of the corresponding node. Next, we let the ``activated'' qutrit interact with the data qubit. An exchange interaction can be implemented on the upper two levels ($|L/R\rangle$) of the address qutrit with two levels of the data qubit. No exchange will occur when the address qutrit is at $|W\rangle$. So that we have
$$|L,0\rangle \mapsto |L,0\rangle,$$
and
$$|L,1\rangle \mapsto |R,0\rangle.$$

\subsubsection{Address bus input}
The address bus input $A(i)$ is to copy the digit $i$ of the address bus to the data qubit of the root node, which is simple CNOT. The address bus input is identical in both qutrit-based and qubit-based schemes. 

Note that this CNOT can also be replaced by a SWAP freely. The only difference could be that a SWAP may require 3 CNOT operations to implement if the CNOT is the basic gate.

\subsubsection{Data bus input}
The data bus input $D_{t/b}(i)$ is to move a digit in the data bus to the data qubit of the root node. This is similar to the address bus input, but the data bus input always requires a SWAP. The reason is further discussed in Supplementary Information Sec.~\ref{Appendix:unitaryQRAM}.

The data bus input of the qutrit-based scheme is a simple SWAP. For the qubit-based scheme, we have to first perform a Hadamard gate to the qubit, then perform this SWAP. This Hadamard changes the basis from $\{|0\rangle, |1\rangle\}$ to $\{|+\rangle, |-\rangle\}$ to allow the data copy operation to distinguish the $|0\rangle$ of the active node from the inactive node. The next part (data copy) will provide more details about this basis change.

\subsubsection{Data copy}
The data copy operation writes the classical data into the QRAM tree leaves to allow it to be moved up to the bus. To write the classical data into qubits, one needs to the classical controlled operation. The data copy of the digit $i$ is represented by $M_{t/b}(i)$.

For the qutrit-based scheme, we use a classical-controlled CNOT operation to implement the data copy, that is
\begin{equation}
M_t = (m_{2p} \textsc{CopyL}_{1,2} + \bar{m}_{2p} I_{1,2} ) (m_{2p+1} \textsc{CopyR}_{1,2}  + \bar{m}_{2p+1} I_{1,2} ).
\end{equation}
Here $\textsc{CopyL} = |W\rangle\langle W| + |L\rangle\langle L\rangle_1 \otimes X_2+ |R\rangle\langle R|_1\otimes I_2$, $\textsc{CopyR} = |W\rangle\langle W| + |L\rangle\langle L|_1\otimes I_2 + |R\rangle\langle R\rangle_1 \otimes X_2$, subscript 1 corresponds to $|a_{l,p}\rangle$ and 2 corresponds to $|d_{l,p}\rangle$. $\bar{m}$ means the flip of $m$.

For the qubit-based scheme, the data copy is implemented by classical-controlled CZ, that is
\begin{equation}
M_b =  (m_{2p} \textsc{CZ}_{1,2} + \bar{m}_{2p} I_{1,2} ) (m_{2p+1} X_1\textsc{CZ}_{1,2}X_1  + \bar{m}_{2p+1} I_{1,2} )
\end{equation}
Here $\textsc{CZ}$ will cause a phase flip on the data qubit when it is at $\{|+\rangle, |-\rangle\}$ basis, and have no effect when at $\{|0\rangle, |1\rangle\}$ basis. For the qubit-based scheme, the data bus input will send the $\{|+\rangle, |-\rangle\}$ state to its corresponding tree leaf, and let other leaves remain in $|0\rangle$ basis. Then this $\textsc{CZ}$ can distinguish the correct data qubit without changing the state of others. Suppose we only send the $\{|0\rangle, |1\rangle\}$ basis encoded data qubits to the leaf, it cannot be separated from the data qubit state that is originally $|0\rangle$. 

Note that there is also another version of the data copy operation, where the last layer of the QRAM tree does not hold address information. This can be understood by adding another layer where each node only has a data qubit. Then in this layer, the data copy is not CNOT or CZ but changes to classical-controlled $X$ and $Z$. These two designs are equivalent, and selecting which design will depend on the actual physical implementation.

\subsection{Layered operation}
The routing, internal swap, and data copy operations will be assembled in layered operations: it is not only performed on a certain node in the QRAM tree, but also on an entire layer. A layered operation can be performed in the same time step, where a node does not share the same qubit as other nodes in the same layer. Layered operations are a manifestation of circuit-level parallelism, which is the origin of the linear time complexity when querying an exponential number of classical data.

The notations of the layered routing and internal swap are $R^{l}$, routing from layer $l$ to layer $l+1$; $I^{l}$, internal swap of layer $l$, respectively. The layered data copy operation directly uses $M$ as the notation.

\subsection{The passiveness of QRAM}

The passiveness and activeness of the QRAM should be carefully treated when discussing a practical QRAM. The layered operation involves exponential gates, and simultaneously applying these gates is an essential requirement of a practical QRAM. In the active QRAM, those gates are compiled independently, which means that they require exponential computing resources. However, with the same scaling of resources, one would have a more efficient classical algorithm, which negates the speedup of the quantum algorithm. On the contrary, a passive QRAM does not require extra resources to compile, which preserves the speedup of the quantum algorithm. The passiveness of the QRAM is an essential requirement for making a practical QRAM, where those layered operations in the QRAM query process should be implemented with a constant or linear number of classical control resources.

There are two possible implementations of a passive QRAM. The first is to simultaneously control these layered operations. We can build a layered QRAM chip where each layer holds a whole QRAM tree layer, and this layer shares the same control resource. This layered QRAM chip requires linear control resources and can be viewed as a passive QRAM. Possible physical systems include the semiconductor or superconducting quantum chip. The second is to use flying qubits, where the data qubits are moved across the layers. The flying-qubit method typically requires no extra resources, including the integrated photonic quantum chip or quantum transmission lines.

\subsection{The unitarity of QRAM}\label{Appendix:unitaryQRAM}

One of the main features of a QRAM is that QRAM can unitarily implement
\begin{equation}
|a_i\rangle|d_i\rangle \mapsto |a_i\rangle|d_i\oplus m_i\rangle,
\end{equation}
instead of the simple
\begin{equation}
|a_i\rangle|0\rangle \mapsto |a_i\rangle|d_i\rangle.
\end{equation}
This feature is required commonly in quantum algorithms when some uncomputing operations are involved. To implement a standard version of the QRAM, one has to first move the data qubit from the bus to the certain tree leaf, perform the $|d_i\rangle$ to $|d_i\otimes m_i\rangle$ mapping, and finally move it back to the bus. As for the simple version, the first ``downward'' step can be ignored, where the input is guaranteed $|0\rangle$. Although we can implement the standard version with two queries of the simple version, that is
\begin{equation}
\begin{aligned}
    &|a_i\rangle|d_i\rangle \\
    \Longrightarrow &|a_i\rangle|d_i\rangle|m_i\rangle \\
    \Longrightarrow &|a_i\rangle|d_i\oplus m_i\rangle|m_i\rangle \\
    \Longrightarrow &|a_i\rangle|d_i\oplus m_i\rangle,
\end{aligned}    
\end{equation}
the fidelity performance does not have any benefit when each QRAM query is erroneous as the main text has assumed.

As the ``top-bottom-top'' route of a data qubit is required from the standard QRAM, the data bus input operation is thus a SWAP (move) instead of a CNOT (copy). If the data bus input is a copy, we will obtain
\begin{equation}
\begin{aligned}
    &|a_i\rangle|d_i\rangle \\
    \Longrightarrow &|a_i\rangle|d_i\rangle|d_i\oplus m_i\rangle,
\end{aligned}    
\end{equation}
where the original $|d_i\rangle$ input is not correctly uncomputed. 

\subsection{Unidirectional and bidirectional routing}

In this section, we discuss the difference between unidirectional routing and bidirectional routing in the QRAM protocol. Unidirectional routing refers to the movement of data solely from the parent to the child, or vice versa. This implies an assumption that the input of one side is always set at $|0\rangle$. On the contrary, bidirectional routing involves a swap between the parent and child, without the aforementioned assumption. In the previous works, only nonparallel protocol was discussed, so that only unidirectional routing is frequently discussed. However, the parallel protocol proposed in this paper emphasizes the importance of the bidirectional routing: the bidirectional routing enables the movement path of each data qubit to be independent from each other, which results in a better efficiency of the parallel protocol than the nonparallel protocol. 

For the unidirectional routing case, firstly we consider that the direction is from parent to child (both children are initialized to $|0\rangle$). So the input is 
$$
|a\rangle|d\rangle|0\rangle|0\rangle,
$$
where the four registers represent the address qudit and the data qubit in $(l,p)$ node, and the data qubits in $(l+1, 2p)$ and $(l+1, 2p+1)$ nodes. $|a\rangle$ and $|d\rangle$ are arbitrary inputs. Then we only have to implement these two mappings:
$$
|L\rangle|d\rangle|0\rangle|0\rangle \mapsto |L\rangle|0\rangle|d\rangle|0\rangle,
$$
and
$$
|R\rangle|d\rangle|0\rangle|0\rangle \mapsto |R\rangle|0\rangle|0\rangle|d\rangle.
$$
Let $L=0$ and $R=1$, the mapping can be represented by the following boolean formula
\begin{equation}
\begin{aligned}
|a\rangle|d\rangle|0\rangle|0\rangle &\mapsto |a\rangle|0\rangle|ad\rangle|\bar{a}d\rangle\\
&=|a\rangle|0\rangle|ad\rangle|ad\otimes d\rangle.
\end{aligned}
\end{equation}
Originally, the implementations of the routing should be two controlled SWAP operations. When the inputs of the children nodes are fixed to $|0\rangle$, it can be implemented by a SWAP followed by a controlled SWAP, that is
\begin{equation}
\begin{aligned}
&\textsc{CSwap}^*\cdot\textsc{SwapR}|a\rangle|d\rangle|0\rangle|0\rangle\\
=&\textsc{CSwap}^*|a\rangle|0\rangle|0\rangle|d\rangle\\
=&|a\rangle|0\rangle|ad\rangle|ad\oplus d\rangle.
\end{aligned}
\end{equation}
Here $\textsc{CSwap}^*$ is a controlled SWAP of the two children's data qubits conditioned by the parent's address qudit, where $\textsc{CSwap}^*|a\rangle|0\rangle|d\rangle = |a\rangle|ad\rangle|ad\oplus d\rangle$. This optimization can reduce a half number of three-qubit gates to two-qubit gates, typically the former being harder to be implemented. For the reverse case (from the child to the parent), we also have to invert the sequence of this gate decomposition.

For the bidirectional routing case, this optimization cannot be applied: two controlled SWAPs are still required. Hereinafter, the parent-to-child routing is represented by $R^{\downarrow}$, the child-to-parent routing is $R^{\uparrow}$, and bidirectional routing is $R^{\updownarrow}$.

In the QRAM query process, positions of $R^{\downarrow}$, $R^{\uparrow}$, and $R^{\updownarrow}$ can be statically identified when the address size $n$ and word length $k$ are given. Therefore, this optimization can be realized without consuming an extra compilation process, which preserves the passiveness of the QRAM.

\subsection{Formal description of the nonparallel protocol}

In this section, the nonparallel protocol of the $(n,k)$-QRAM is formally described. The three phases are address setting, data copy and uncomputing.

The address setting phase will copy every address qubit to the corresponding layer, that is
\begin{equation}
\textsc{AddressSetting}_i = I_{i}\left(\prod_{l=0}^{i-1}R^{\downarrow}_i\right)A(i).
\end{equation}
Then the whole address setting phase is
\begin{equation}
\textsc{AddressSetting} = \prod_{i=0}^{n-1}\textsc{AddressSetting}_i.
\end{equation}

The data copy phase moves the data to the leaf, performs the data copy operation, then moves back to the root, that is
\begin{equation}
    \textsc{DataFetch}_i = D(i)\left(\prod_{l=0}^{n-1}R^{\downarrow}_{i}\right)^{\dagger}M_i\left(\prod_{l=0}^{n-1}R^{\downarrow}_i\right)D(i).
\end{equation}
Because we have $R^{\downarrow} = (R^{\uparrow})^{\dagger}$, we can also write
\begin{equation}
    \textsc{DataFetch}_i = D(i)\left(\prod_{l=0}^{n-1}R^{\uparrow}_{n-1-i}\right)M_i\left(\prod_{l=0}^{n-1}R^{\downarrow}_i\right)D(i).
\end{equation}
The data copy for all $k$ data qubits are
\begin{equation}
\textsc{DataFetch} = \prod_{i=0}^{k-1}\textsc{DataFetch}_i.
\end{equation}
The uncomputing phase is the reverse of the address setting phase, that is
\begin{equation}
\textsc{Uncomputing}=\textsc{AddressSetting}^\dagger.
\end{equation}
Similarly, we can directly write
\begin{equation}
\textsc{Uncomputing}=\prod_{i=0}^{n-1}\textsc{Uncomputing}_{n-1-i},
\end{equation}
and 
\begin{equation}
\textsc{Uncomputing}_i = A_{i}\left(\prod_{l=0}^{i-1}R^{\uparrow}_{l-i-1}\right)I_{i+1}^{\dagger}.
\end{equation}
Finally we have
\begin{equation}
    U_{\textsc{Nonparallel}} = \textsc{Uncomputing}\cdot\textsc{DataFetch}\cdot\textsc{AddressSetting}.
\end{equation}

Note that $U_{\textsc{Nonparallel}}$ is also a general description to $(n,k)$-QRAM, where 
$U_{\textsc{Nonparallel}}|i\rangle|z\rangle = |i\rangle|z\oplus m[i]\rangle$.

\subsection{Formal description of the parallel protocol}

The parallel protocol considers circuit parallelism and a compact data access sequence. To formally describe the parallel protocol, we focus on extracting the time sequence of operations. First we individually show the time sequence of each phase.


\paragraph{Address setting phase}
In the address setting phase, we can simply advance every operation as early as possible to obtain a compact time step by applying the commutation rules:
\begin{equation}
\begin{aligned}
    \left[A(i), I_j\right] &= \delta_j, \\
    \left[A(i), R_j\right] &= \delta_j, \\
    \left[I_i, R_j\right] &= \delta_{i,j} + \delta_{i+1,j}. \\
\end{aligned}
\end{equation}
The original operation sequence is
\begin{equation}
\begin{aligned}
    &A(0) \rightarrow I_0  \\
   \rightarrow &A(1) \rightarrow R_0 \rightarrow I_1 \\
   \rightarrow &A(2) \rightarrow R_0 \rightarrow R_1 \rightarrow I_2 \\
   \rightarrow &A(3) \rightarrow R_0 \rightarrow R_1 \rightarrow R_2 \rightarrow I_3 \\
   \rightarrow &A(4) \rightarrow R_0 \rightarrow R_1 \rightarrow R_2 \rightarrow R_3 \rightarrow I_4 \\
\end{aligned}
\end{equation}
Then we can rewrite the sequence by rearranging these operations, where
\begin{equation}
\begin{aligned}
    &A(0) \rightarrow I_0  \\
   \rightarrow &A(1) \rightarrow R_0 \rightarrow \textcolor{red}{A(2)} \rightarrow I_1 \\
   \rightarrow &R_0 \rightarrow \textcolor{red}{A(3)} \rightarrow R_1 \rightarrow \textcolor{red}{R_0} \rightarrow I_2 \rightarrow \textcolor{red}{A(4)}\\
   \rightarrow &R_1 \rightarrow \textcolor{red}{R_0} \rightarrow  R_2 \rightarrow \textcolor{red}{R_1} \rightarrow I_3 \\
   \rightarrow & R_2 \rightarrow R_3 \rightarrow I_4
\end{aligned}
\end{equation}
The red-color operations are the operations moved forward by applying the commutation rules.

\paragraph{Data fetch phase}
For the data fetch phase, the data bit $i$ fetches its corresponding memory by applying the unitary 
\begin{equation}
    \textsc{DataFetch}_i = D(i)\left(\prod_{l=0}^{n-1}R^{\uparrow}_{n-1-i}\right)M_i\left(\prod_{l=0}^{n-1}R^{\downarrow}_i\right)D(i).
\end{equation}
We can write the operation sequence of the data fetch phase as
\begin{equation}
\begin{aligned}
    &D(0) \rightarrow R_0^\downarrow \rightarrow R_1^\downarrow \rightarrow ... \rightarrow R_{n-1}^\downarrow  \rightarrow M(0) \rightarrow R_{n-1}^\uparrow \rightarrow R_{n-2}^\uparrow \rightarrow... \rightarrow R_2^\uparrow \rightarrow R_1^\uparrow \rightarrow R_0^\uparrow \rightarrow D(0) \\
    &D(1) \rightarrow R_0^\downarrow \rightarrow R_1^\downarrow \rightarrow ... \rightarrow R_{n-1}^\downarrow  \rightarrow M(1) \rightarrow R_{n-1}^\uparrow \rightarrow R_{n-2}^\uparrow \rightarrow... \rightarrow R_2^\uparrow \rightarrow R_1^\uparrow \rightarrow R_0^\uparrow \rightarrow D(1) \\
    &D(2) \rightarrow R_0^\downarrow \rightarrow R_1^\downarrow \rightarrow ... \rightarrow R_{n-1}^\downarrow  \rightarrow M(2) \rightarrow R_{n-1}^\uparrow \rightarrow R_{n-2}^\uparrow \rightarrow... \rightarrow R_2^\uparrow \rightarrow R_1^\uparrow \rightarrow R_0^\uparrow \rightarrow D(2) ...
\end{aligned}
\end{equation}
To allow circuit-level parallelism, we may try to let every digit starts after two operations, where
\begin{equation}
\begin{aligned}
    &D(0) \rightarrow R_0^\downarrow \rightarrow R_1^\downarrow \rightarrow ... \rightarrow R_{n-1}^\downarrow  \rightarrow M(0) \rightarrow R_{n-1}^\uparrow \rightarrow R_{n-2}^\uparrow \rightarrow... \rightarrow R_2^\uparrow \rightarrow R_1^\uparrow \rightarrow R_0^\uparrow \rightarrow D(0) \\
    &I \rightarrow I \rightarrow D(1) \rightarrow R_0^\downarrow \rightarrow R_1^\downarrow \rightarrow ... \rightarrow R_{n-1}^\downarrow  \rightarrow M(1) \rightarrow R_{n-1}^\uparrow \rightarrow R_{n-2}^\uparrow \rightarrow... \rightarrow R_2^\uparrow \rightarrow R_1^\uparrow \rightarrow R_0^\uparrow \rightarrow D(1) \\
    &I \rightarrow I \rightarrow I \rightarrow I \rightarrow D(2) \rightarrow R_0^\downarrow \rightarrow R_1^\downarrow \rightarrow ... \rightarrow R_{n-1}^\downarrow  \rightarrow M(2) \rightarrow R_{n-1}^\uparrow \rightarrow R_{n-2}^\uparrow \rightarrow... \rightarrow R_2^\uparrow \rightarrow R_1^\uparrow \rightarrow R_0^\uparrow \rightarrow D(2) ...
\end{aligned}
\end{equation}
Each operation corresponds to a single time step, and operations on different lines are intended to be executed concurrently. The majority of these operations do not overlap in the same layer, allowing for simultaneous execution. There are a few instances where operations share the same layer. Nevertheless, due to their placement in the time sequence, certain operations like $R^\uparrow_k$ and $R^\downarrow_k$ always occur within the same time step. These paired operations can be combined into a single operation, utilizing bidirectional routing as a replacement. In other words, $R^\uparrow_kR^\downarrow_k$ can be substituted with $R_k^\updownarrow$.



\subsection{Proof of time complexity}
The time complexity of the nonparallel and parallel protocols can be directly extracted from the formal descriptions shown in the above section. Treating each fundamental operation as one time step, we can obtain the total number of time steps for each phase.

For the address setting phase, the address setting phase is $3n-1$ (the circuit-level parallelism can be both applied to the parallel and nonparallel protocols). The time complexity is $O(n)$.

For the data fetch phase, the number of time steps for the nonparallel protocol is $O(k(n+1))$, where each digit uses $n+1$ steps and $k$ digits in total. No operations can be executed simultaneously. The number of time steps for the parallel protocol is $O(2k+n-1)$, where the first digit uses $n+1$ steps, and one more digit consumes extra 2 steps. The time complexity is for the two protocols are thus $O(nk)$ and $O(n+k)$, respectively.

\section{Proof of error scaling\label{Appendix:error}}

\subsection{A review of proof for error-resilience of bucket-brigade QRAM}
 To begin with, we review the idea of proof for noise-resilience of the QRAM in \cite{ErrorResilience}. For a QRAM query instance
 $$
 U_{\mathrm{QRAM}}\sum_i\alpha_i|a_i\rangle(\sum_j|z_j\rangle)= \sum_i\alpha_i|a_i\rangle(\sum_j|z_j\oplus m_i\rangle).
 $$
Each $|a_i\rangle$ is called a ``branch''. First, we classify all branches into good branches and bad branches for a certain error configuration $c$, which represents a configuration of Kraus operators on all qudits and all time steps. A good branch means that the state of all qudits that are related to the corresponding address remains unchanged after applications of Kraus operators. The overlap between good branches and the ideal final state satisfies
\begin{equation}
    \langle\psi_{\mathrm{out}} | \mathrm{good}(c)\rangle = \Lambda(c),
\end{equation}
where $\Lambda(c)$ is the sum of the amplitude of good branches $\Lambda(c)=\sum_{i\in \mathrm{good}}|\alpha_i|^2$. The fidelity of error configuration $c$ satisfies $F(c)\geq (2\Lambda(c)-1)^2$ for $\Lambda(c)>1/2$. 

The fidelity is the average of $F(c)$ for all possible error configurations, that is
\begin{equation}\label{eqn:fidelity1}
    \begin{aligned}
    F &= \mathbb{E}(F(c)) \geq \mathbb{E}(2\Lambda(c)-1)^2\\
    &\geq (2\mathbb{E}(\Lambda)-1)^2.
    \end{aligned}
\end{equation}
Here $\mathbb{E}(\Lambda)$ is the average fraction of good branches, which is computed iteratively and finally yields $\mathbb{E}(\Lambda) = (1-\epsilon)^{T\log N} \geq 1-\epsilon T\log N$. Combining this inequality with Eqn.~(\ref{eqn:fidelity1}), we obtain the final result
\begin{equation}
    F \geq 1 - 4\epsilon T\log N = 1-4\epsilon nT.
\end{equation}
For the qubit-based scheme, most of the processes preserve the same, except $\mathbb{E}(\Lambda)_b = (1-\epsilon)^{Tn^2} \geq 1-\epsilon Tn^2$.

\subsection{Error scalings for different protocols}

In the main text, we stated that the error scaling of each protocol can be obtained by replacing the time complexity $T$ and the error of each qubit $\epsilon$ with respective values. To show this can be directly used to extend the proof for the parallel protocol, we should also show that the ``good branch'' has the same definition in these protocols.

First, the error in the bad branch will not be propagated to any good branch. At the data fetch phase, the data bits are only transferred through the data qubit in the node. When there is no error on the address qubit, a bad branch cannot pass the error to the good branch the same as how the normal bucket-brigade architecture does. Second, the fraction of the good branch is $1-O(nT\epsilon)$. Because $T\epsilon$ is the total error for a qubit that undergoes $T$ time steps, we only have to change $T$ to the number of time steps in the parallel protocol.

For the nonparallel protocol, it repeats the data fetch phase $k$ times. Note that the data fetch phase still has $O(n^2\epsilon)$ error scaling even if there is no error in the address setting and uncomputing phase, which has been proved in \cite{ErrorResilience}. Therefore, the error scaling for the nonparallel protocol is at least $O(kn^2\epsilon)$ when $k$ scales. 

The high-bandwidth protocol has the same structure and process as the original bucket-brigade QRAM. We can bind all data qubits in a node as a quasiparticle with $2^k$ levels. When each qubit has an error rate $\epsilon$, this quasiparticle's fidelity is $(1-\epsilon)^k \sim 1-k\epsilon$. Therefore, we only have to replace $\epsilon$ with $k\epsilon$ in the error scaling expression.


\section{Comparison with the quantum-walk-based architecture}
Recently, a quantum-walk-based QRAM architecture has been proposed \cite{QWQRAM,QWQRAM2}. This architecture also considered the case where the word length is more than 1. In the proposal, an $(n,k)$-QRAM can be implemented using $n+k$ quantum walkers on a directed graph, which uses $O(n+k)$ qubits and achieves $O(n\log (n+k))$ time complexity.

The scaling of the qubit number and time does not mean it does not consume exponential resources. Instead, a quantum walker is like a ``flying qubit'' and can travel coherently through quantum switches. Therefore, the number of quantum switches and the circuit width is exponential to the address length. 

The time complexity of the quantum-walk QRAM is $O(n\log(n+k))$. Intuitively, when $k$ is a constant (such as $k=1$), an $O(n\log n)$ is already a slow-down over the bucket-brigade architecture, which is $O(n)$; also, the quantum-walk QRAM has a speedup when $k$ is sufficiently larger than $k$ to let $n+k > n\log(n+k)$. While this work does not explicitly present the error scaling, we would not compare it directly with our protocols. However, as this method is likely to apply $k$ qubits to independently query each bit of the data, this method is similar to the high-bandwidth protocol with bandwidth $k$, where the error rate would be at least $O(n^2k)$. 

\section{The error filtration of QRAM and the cost factor}

The error filtration method can suppress the error of an arbitrary-size black box quantum operation. As have discussed in the main text, the QRAM is not likely to be completely built with logical qubits, this method is likely to be applied to the QRAM with noisy physical qubits to reduce the error.

Suppose we have a noisy quantum process $\mathcal{E}$ which corresponds to a perfect unitary $U$. Then given arbitrary input $|\psi\rangle$, the infidelity between $\rho = \mathcal{E}(|\psi\rangle\langle \psi|)$ and $\rho_{\mathrm{perfect}} = U|\psi\rangle\langle \psi|U^\dagger$ is scales as $\epsilon$. Then the error filtration can use $T$ queries to this process $\mathcal{E}$ to obtain a state that has infidelity scales $\epsilon/T$. Note that the error filtration requires $\log T$ number of perfect ancilla qubits. Thus we can apply the error filtration with logical qubits to filter the noisy process in the bus of the QRAM.

The error filtration method offers us a trade-off between the error rate and the query time of the QRAM. When the error is reduced to $1/T$, the query time is raised to $T$ times. From this point, we propose the cost factor in the main text, which multiplies the time complexity by the error rate. Despite how to apply the error filtration, the cost factor can be a general measure of the costs of different QRAM protocols.

\end{document}